\documentclass[aps,prb,reprint,superscriptaddress,longbibliography]{revtex4-2}

\usepackage{graphicx}
\usepackage{amsmath}
\usepackage{bm}
\usepackage[dvipsnames]{xcolor}

\newcommand{\taufast}{\tau_{\mathrm{fast}}}
\newcommand{\tauslow}{\tau_{\mathrm{slow}}}
\newcommand{\gammafast}{\gamma_{\mathrm{fast}}}
\newcommand{\gammaslow}{\gamma_{\mathrm{slow}}}
\newcommand{\taus}{\tau_{\mathrm{s}}}
\newcommand{\Tg}{T_{\mathrm{g}}}
\newcommand{\PNCP}{Pd$_{42.5}$Ni$_{7.5}$Cu$_{30}$P$_{20}$}

\begin{document}

\title{Nanoscale two-component structural relaxation in the fragile metallic glass \PNCP\ revealed by five-dimensional scanning transmission electron microscopy}

\author{Katsuaki Nakazawa}
\affiliation{Center for Basic Research on Materials, National Institute for Materials Science,
1-2-1 Sengen, Tsukuba, Ibaraki 305-0047, Japan}
\author{Hidemi Kato}
\author{Tetsu Ichitsubo}
\affiliation{Institute for Materials Research, Tohoku University, 2-1-1 Aoba-ku, Sendai, Miyagi 980-8577, Japan}
\author{Shinji Kohara}
\author{Kazutaka Mitsuishi}
\affiliation{Center for Basic Research on Materials, National Institute for Materials Science,
1-2-1 Sengen, Tsukuba, Ibaraki 305-0047, Japan}
\author{Koichi Tsuchiya}
\affiliation{Research Center for Structural Materials, National Institute for Materials Science,
Tsukuba, Ibaraki 305-0047, Japan}

\date{\today}

\begin{abstract}
Measurements without nanoscale spatial resolution have shown that the structural relaxation of metallic glasses can consist of multiple components, but it remains unclear whether these components coexist locally or appear only
after averaging over regions that each relax with one component. Using in situ five-dimensional scanning transmission electron
microscopy with a sub-nanometer probe, we measured the local relaxation of the fragile metallic glass
\PNCP\ near its glass transition temperature and tested at each position whether one or two
Kohlrausch--Williams--Watts components are required. At 558~K most positions require two components,
so the relaxation within nanometer-scale volumes consists of two distinct processes rather than
one process. As the temperature is increased toward the glass transition, the fraction of positions requiring two components decreases,
and the local relaxation changes from two-component to one-component. Positions with a higher degree
of structural order show a larger contribution from the slow component, which indicates that the two-component
relaxation is linked to the local atomic structure.
\end{abstract}

\maketitle

\section{Introduction}

Relaxation is a central issue in glass transition phenomena. One of the main features of the glass
transition is the rapid increase in relaxation time (or viscosity) near the glass transition temperature
($\Tg$), and the steepness of this increase is characterized by the
fragility~\cite{Ediger1996,Berthier2011,Angell1988,Angell1995,Debenedetti2001}. Another feature of
glassy dynamics is that the structural relaxation is non-exponential; it is commonly described by the
Kohlrausch--Williams--Watts (KWW) function, $\exp[-2(t/\tau)^{\gamma}]$, with a stretching exponent
$\gamma<1$. These two features are closely related: glass formers with higher fragility tend to exhibit
more strongly non-exponential relaxation~\cite{Bohmer1993}. Understanding the origin of the
non-exponential relaxation is therefore essential for understanding the glass transition, particularly
in fragile glass formers. Whether the stretched decay reflects an overlap of regions with
different relaxation times or a relaxation that is non-exponential at every location cannot be decided
from spatially averaged measurements~\cite{Ediger2000,Richert2002}.

In metallic glasses, recent measurements have shown that the relaxation on the time scale of the
structural relaxation cannot always be described by one process. X-ray photon correlation
spectroscopy (XPCS) has revealed compressed exponential decays, intermittent aging, and decays with
multiple components in the intensity correlation functions~\cite{Luo2017,Ruta2012,Evenson2015}. Dynamic mechanical
analysis has also shown that the structural relaxation proceeds with two components in metallic
glasses~\cite{Okumura1992}. However, these techniques average over the spatial distribution of the relaxation. It therefore remains unclear whether the multi-component relaxation
occurs locally at each nanoscale region, or whether it appears only after averaging over regions that
each relax with one component.

Since relaxation proceeds through rearrangements of atoms and their neighbors, it is expected to be
closely linked to the local atomic structure. Both experiments and simulations have shown that the
dynamics is spatially heterogeneous and correlated with the
structure~\cite{Tong2018,Ichitsubo2005,Ichitsubo2011}. In particular, fragile metallic glasses have been
suggested to consist of strongly bonded regions and weakly bonded regions with different
mobilities~\cite{Ichitsubo2005,Ichitsubo2011}. Electron correlation microscopy and 5D-STEM have recently
enabled direct real-space mapping of the relaxation time with nanometer resolution, revealing
domain-like distributions of relaxation times and their correlation with local
structure~\cite{Spangenberg2021,Zhang2018,Nakazawa2024}. In these studies, however, the local relaxation
was analyzed with a one-component KWW function, and the possibility that multiple relaxation processes coexist
at the same location has not been examined.

In this study, we address two questions: (i) whether the multi-component relaxation occurs at the
nanometer scale, and (ii) if so, what local structure determines the relative contributions of the fast
and slow components. We investigated a fragile metallic glass, in which non-exponential relaxation is expected to be
pronounced. We employed the 5D-STEM method, which uses a high-speed pixelated STEM detector to record
the spatiotemporal distribution of diffraction patterns~\cite{Nakazawa2023,Huang2024}. Although 5D-STEM
offers lower temporal resolution than XPCS, it enables spatially resolved measurements and simultaneous
analysis of structure and dynamics at each probe position. We fitted the local correlation functions
with both a one-component KWW function and a sum of two KWW functions, and determined by statistical model
selection whether two components are required at each position. We then mapped the relaxation times of
the fast and slow components and their fraction near $\Tg$, and correlated them with structural
parameters extracted from the same diffraction patterns.

\section{Methods}

The sample was the metallic glass \PNCP, which is fragile with a fragility index of approximately
58.5~\cite{Kato2006,Ichitsubo2011} and has $\Tg \approx 570$~K~\cite{Haruyama2011}. The TEM sample was prepared by focused ion beam milling; the thickness of the sample was
50--60~nm. A JEM-ARM200F (JEOL Ltd.) equipped with a cold field emission gun and a 4DCanvas camera (JEOL
Ltd.) was used. The acceleration voltage and the convergence semi-angle were set at 200~kV and 1.6~mrad,
and the probe current was 8.9~pA. The corresponding probe size is 0.8~nm in full
width at half maximum (FWHM), and the depth of focus of about 500~nm is much larger than the sample
thickness, so the probe size is essentially constant through the foil. The scan step, the number of scanning points and the dwell time were
0.27~nm, $60 \times 180$ points and 133~$\mu$s. The pixel size and the number of pixels of the
diffraction patterns were 0.058~nm$^{-1}$ and $264 \times 264$. The diffraction patterns were
$4 \times 4$ binned to increase the signal-to-noise ratio, giving a pixel size of 0.23~nm$^{-1}$ and
$66 \times 66$ pixels. The time resolution of the 5D-STEM observation was 1.9~s with this setting.
The 5D-STEM observations were conducted from 558 to 578~K using an in-situ heating holder (Fusion,
Protochips Inc.). Up to 563~K, 600 diffraction patterns were acquired at each scan point, and from
568~K, 300 diffraction patterns were acquired.

\section{Results}

\subsection{Local two-component relaxation}

\begin{figure}[tb]
\includegraphics[width=\columnwidth]{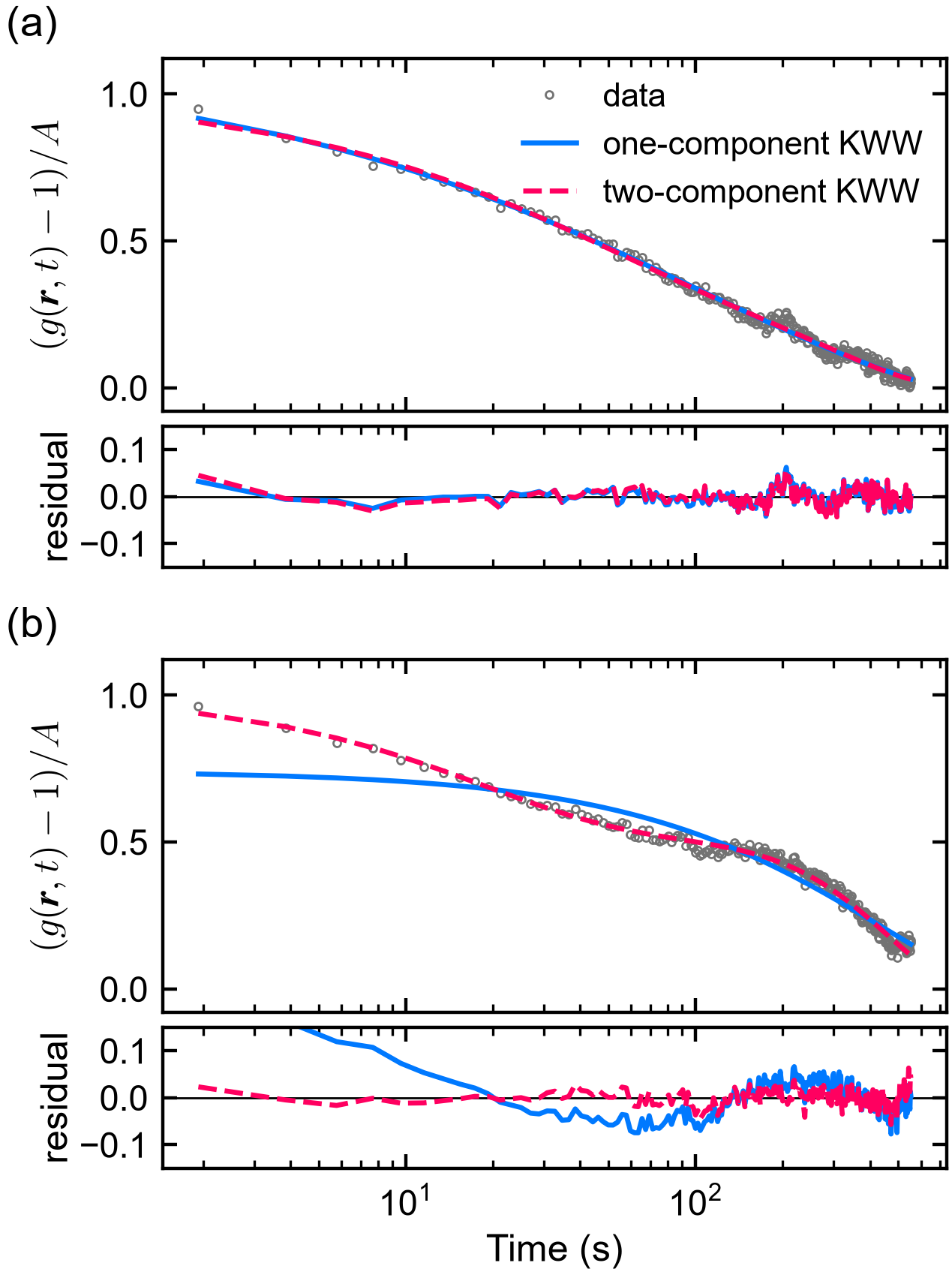}
\caption{Correlation functions $g(\bm{r},t)$ at two positions at 558~K, fitted with a one-component KWW
function (blue solid line) and with the sum of two KWW functions [Eq.~\eqref{eq:double}, red dashed line].
The data and both fits are shown as $(g(\bm{r},t)-1)/A$, with $A$ from the fit of Eq.~\eqref{eq:double}. (a) A position
where a one-component KWW function is sufficient ($\tau = 457$~s, $\gamma = 0.46$). (b) A position where two
components are required ($\taufast = 40$~s, $\tauslow = 637$~s, $f = 0.48$). The lower panels show the
residuals; the one-component KWW function leaves a systematic deviation in (b) but not in (a).}
\label{fig:fit}
\end{figure}

We measured the spatial distribution of the relaxation times from the five-dimensional intensity
distribution $I(\bm{r},\bm{k},t)$ acquired by 5D-STEM, where $\bm{r}$ is the two-dimensional real-space
position, $\bm{k}$ is the two-dimensional reciprocal-space position, and $t$ is time. The relaxation was
evaluated from the decay of the temporal correlation of the time-series diffraction patterns at each
position, using the two-time correlation function (2TCF)~\cite{Madsen2010},
\begin{equation}
G(\bm{r},k,t_1,t_2)=
\frac{\left\langle I(\bm{r},\bm{k},t_1)\,I(\bm{r},\bm{k},t_2)\right\rangle_{\phi}}
     {\left\langle I(\bm{r},\bm{k},t_1)\right\rangle_{\phi}
      \left\langle I(\bm{r},\bm{k},t_2)\right\rangle_{\phi}} .
\label{eq:2tcf}
\end{equation}
Here the subscript $\phi$ indicates that the azimuthal average is calculated on the ensemble of pixels
over a range of spatial frequency. In this experiment, we set the range from 4.4 to 4.6~nm$^{-1}$, where
the peak of the halo pattern appeared. By taking an average over a fixed delay time $t$, the temporal
correlation $g$ was obtained,
\begin{equation}
g(\bm{r},k,t)=\left\langle G(\bm{r},k,t_1,t_1+t)\right\rangle_{t_1} .
\label{eq:g}
\end{equation}
In this article, since we do not focus on the $k$ dependence of $g(\bm{r},k,t)$, we omit
$k$ hereafter. It is worth noting that, unlike the one-time correlation function, the 2TCF
does not force $g(\bm{r},t)$ to converge to zero within the observation time, allowing for the
measurement of longer relaxation processes~\cite{Nakazawa2025}.
Figure~\ref{fig:fit} shows $g(\bm{r},t)$ at 558~K for two representative positions $\bm{r}$.
A one-component KWW function describes the data well in Fig.~\ref{fig:fit}(a) but poorly in Fig.~\ref{fig:fit}(b).
At the position in Fig.~\ref{fig:fit}(b), it cannot reproduce both the rapid decay within the first 10 s and the second decay after about 100 s.
Motivated by this observation and by the structural relaxation with two components reported by mechanical spectroscopy~\cite{Okumura1992}, we introduced
the sum of two KWW functions~\cite{Spangenberg2021,Bikondoa2017},
\begin{equation}
\begin{split}
\frac{g(\bm{r},t)-1}{A}={}&
f\exp\!\left[-2\left(\frac{t}{\taufast}\right)^{\gammafast}\right]\\
&+(1-f)\exp\!\left[-2\left(\frac{t}{\tauslow}\right)^{\gammaslow}\right],
\end{split}
\label{eq:double}
\end{equation}
with $\taufast \le \tauslow$. Here $\taufast$ and $\tauslow$ are the relaxation times of the fast and
slow components, $\gammafast$ and $\gammaslow$ are their stretching exponents, $f$ is the fraction of
the total relaxation contributed by the fast component, and $A$ is a scaling factor.
Equation~\eqref{eq:double} is the minimal extension of the one-component KWW function; we did not consider
three or more components because no behavior beyond two components was observed within the present time
window. The fitting was performed with a trust-region reflective algorithm, with bounds of
$1$--$10^{7}$~s for $\taufast$ and $\tauslow$, $0.2$--$2.0$ for $\gammafast$ and $\gammaslow$, and
$0$--$1$ for $f$. For the one-component KWW function, a small baseline $B$ ($|B|\le 0.1$) was allowed. At high
temperatures the relaxation is completed within the observation time, after which $g(\bm{r},t)$
fluctuates around zero because of noise, and the baseline absorbs this noise floor. No baseline was
included in Eq.~\eqref{eq:double}, because a slow component with $\tauslow$ much longer than the
observation time is indistinguishable from a constant baseline.

\begin{figure}[tb]
\includegraphics[width=\columnwidth]{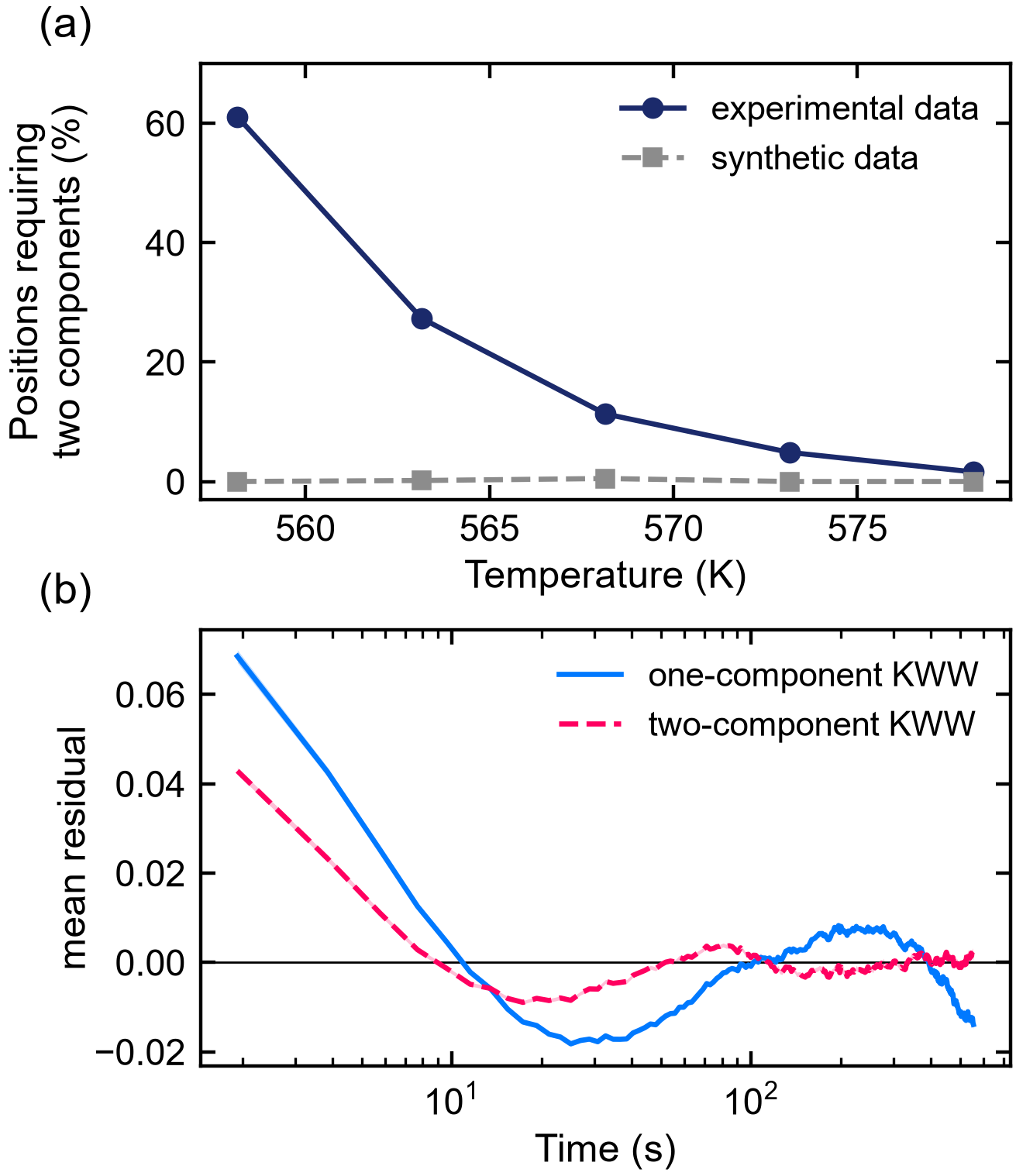}
\caption{Statistical evidence for the two-component relaxation. (a) Fraction of positions in the
experimental data classified as two-component by the autocorrelation-corrected BIC (blue circles). Gray
squares: false-positive rate of the same procedure in a separate control on synthetic one-component KWW
data with noise matching the measured residuals. (b) Mean residual of the fits to the experimental data over
the two-component positions at 558~K for the one-component (blue) and two-component (red) KWW functions.
Shaded bands show the standard error of the mean.}
\label{fig:stat}
\end{figure}

\begin{figure*}[t]
\includegraphics[width=\textwidth]{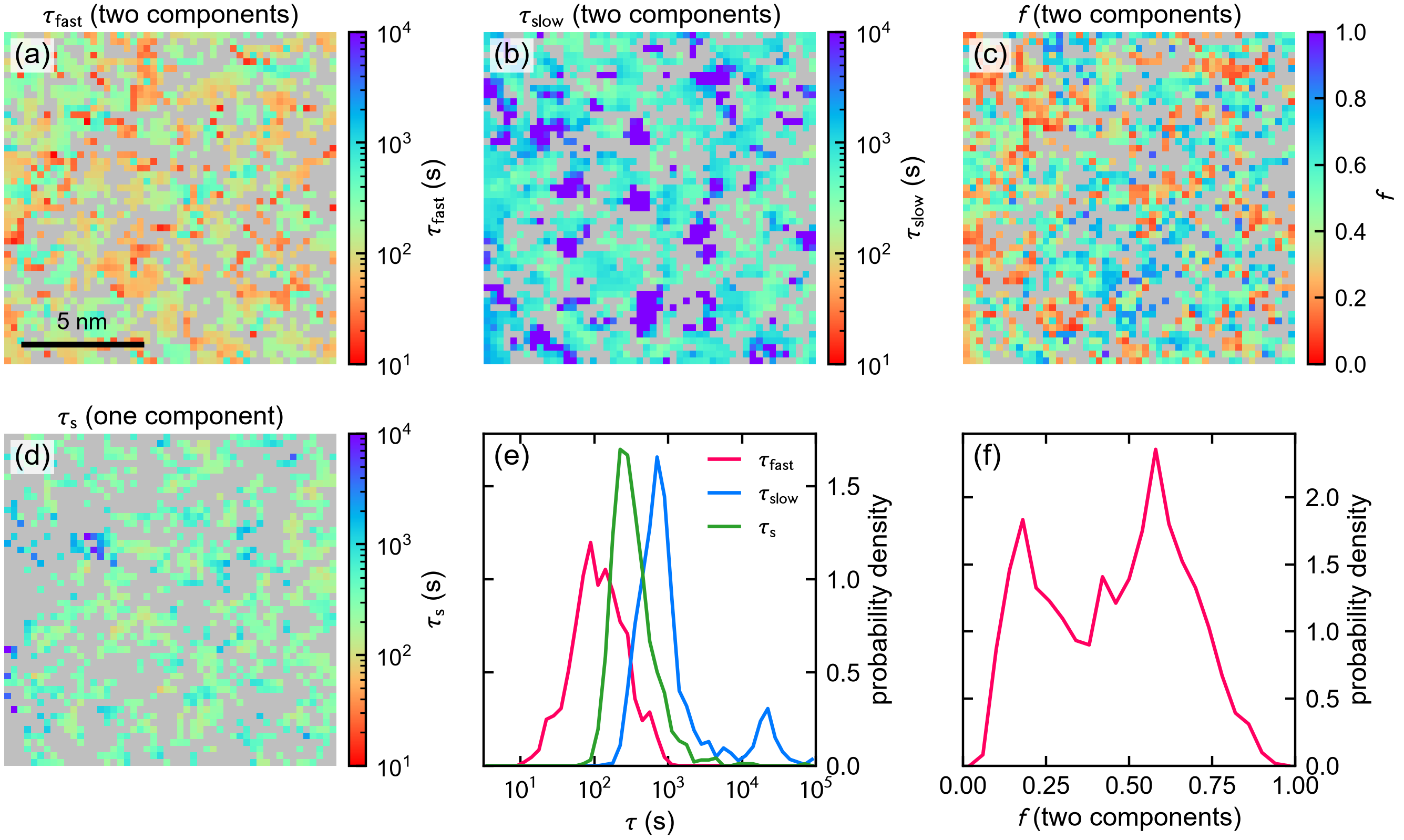}
\caption{Local relaxation parameters at 558~K. Each map shows a parameter only at the positions where the
corresponding model was selected: (a) $\taufast$, (b) $\tauslow$ and (c) $f$ at the two-component positions
and (d) $\taus$ at the one-component positions. The other positions are shown in gray. (e) Probability
densities of $\taufast$, $\tauslow$ (two-component positions) and $\taus$ (one-component positions) and
(f) of $f$ (two-component positions), obtained from normalized histograms (bin widths 0.1 in
$\log_{10}\tau$ and 0.04 in $f$) and drawn as lines. The scale bar is 5~nm.}
\label{fig:maps}
\end{figure*}

Because Eq.~\eqref{eq:double} has more free parameters than the one-component KWW function, a better fit alone
does not demonstrate the presence of two components. We therefore determined at each position whether
the second component is statistically required, using the Bayesian information criterion (BIC) evaluated with an effective number of data points that
accounts for the serial correlation of the fit residuals (Sec.~II of the Supplemental Material~\cite{SM}).\nocite{Egerton2011,Chatterjee2021} A position was
classified as two-component when $\Delta \mathrm{BIC} = \mathrm{BIC}_{\mathrm{two}} -
\mathrm{BIC}_{\mathrm{one}} < 0$; otherwise it was classified as one-component, and its relaxation time
$\taus$ was taken from the one-component KWW fit.

The blue circles in Fig.~\ref{fig:stat}(a) show the fraction of positions in the experimental data
classified as requiring two components at each temperature. At 558~K, 61\% of the positions require the
two-component function even with this conservative criterion.

Because noise can make the BIC favor the two-component function by chance, we estimated the
false-positive rate of this classification in a separate control analysis on synthetic data. We generated
one-component KWW decays, with $\tau$ and $\gamma$ bootstrapped from the one-component positions of the
experimental data, and added noise with the same amplitude and serial correlation as the measured residuals;
by construction, these synthetic data contain no second component. Applying the identical fitting and
selection procedure to them (600 realizations per temperature), the false-positive rate was 0.5\% or less
at all temperatures [gray squares in Fig.~\ref{fig:stat}(a)], more than 100 times smaller than the
fraction found in the experimental data at 558~K. The two-component relaxation at low temperatures therefore
cannot be explained by noise.

The residuals of the fits to the experimental data provide independent evidence [Fig.~\ref{fig:stat}(b)].
Averaged over the two-component positions at 558~K, the one-component KWW fit leaves a systematic,
oscillating residual (positive at short times, negative at intermediate times, and positive again at long
times) that is reproducible over hundreds of positions, whereas the residual of the two-component fit is
flat beyond the first two data points. The root-mean-square of the averaged residual is
1.3--2.2 times larger for the one-component KWW fit than for the two-component fit. At the two shortest delay times
the two-component fit also leaves a positive residual, but excluding these points changes neither the
classification nor the conclusions (Sec.~II of the Supplemental Material); we therefore include
them and treat $\taufast$ as accurate to about 30\%. A nested $F$-test gave consistent results (Table~S1
in the Supplemental Material).

\subsection{Spatial distribution of the two-component relaxation}

Figure~\ref{fig:maps} shows maps of $\taufast$, $\tauslow$ and $f$ at the two-component positions and
of $\taus$ at the one-component positions at 558~K. All parameters are domain-like and spatially
heterogeneous, and the maps differ from each other,
indicating that the speed of relaxation and the relative weight of the fast and slow components vary
independently in space. Among the two-component positions, the nearest-neighbor spatial correlation of
$f$ is 0.40, which is weaker than those of $\log \tauslow$ (0.63) and of the fluctuation (0.83), but
clearly nonzero compared with the value of about zero obtained after randomly shuffling the positions. Although spatial heterogeneity of
one-component relaxation has been reported~\cite{Spangenberg2021,Zhang2018,Nakazawa2024},
Fig.~\ref{fig:maps} directly shows that the two-component relaxation itself is spatially heterogeneous at the
nanometer scale. At the two-component positions the stretching exponents of both components vary widely
from position to position (Fig.~S3 in the Supplemental Material). At the one-component positions the
decay is strongly stretched, with $\gamma$ well below 1.

The reliability of the fitted parameters was examined from the fraction of relaxation completed within
the observation time (Fig.~S3 in the Supplemental Material); when less than half of the relaxation occurs
within the observation time, the error in $\tau$ can exceed 50\%~\cite{Nakazawa2025}. For most positions
more than 90\% of the fast component has relaxed within the observation time, so that $\taufast$ is
reliable. In contrast, at some positions $\tauslow$ exceeds
$10^{5}$~s; there the slow component hardly changes during the measurement and acts as a nearly constant
background. These positions (5\% or less of the two-component positions at each temperature) were excluded
from the averages of $\tauslow$.

\subsection{Temperature dependence and assignment of the two components}

\begin{figure}[tb]
\includegraphics[width=\columnwidth]{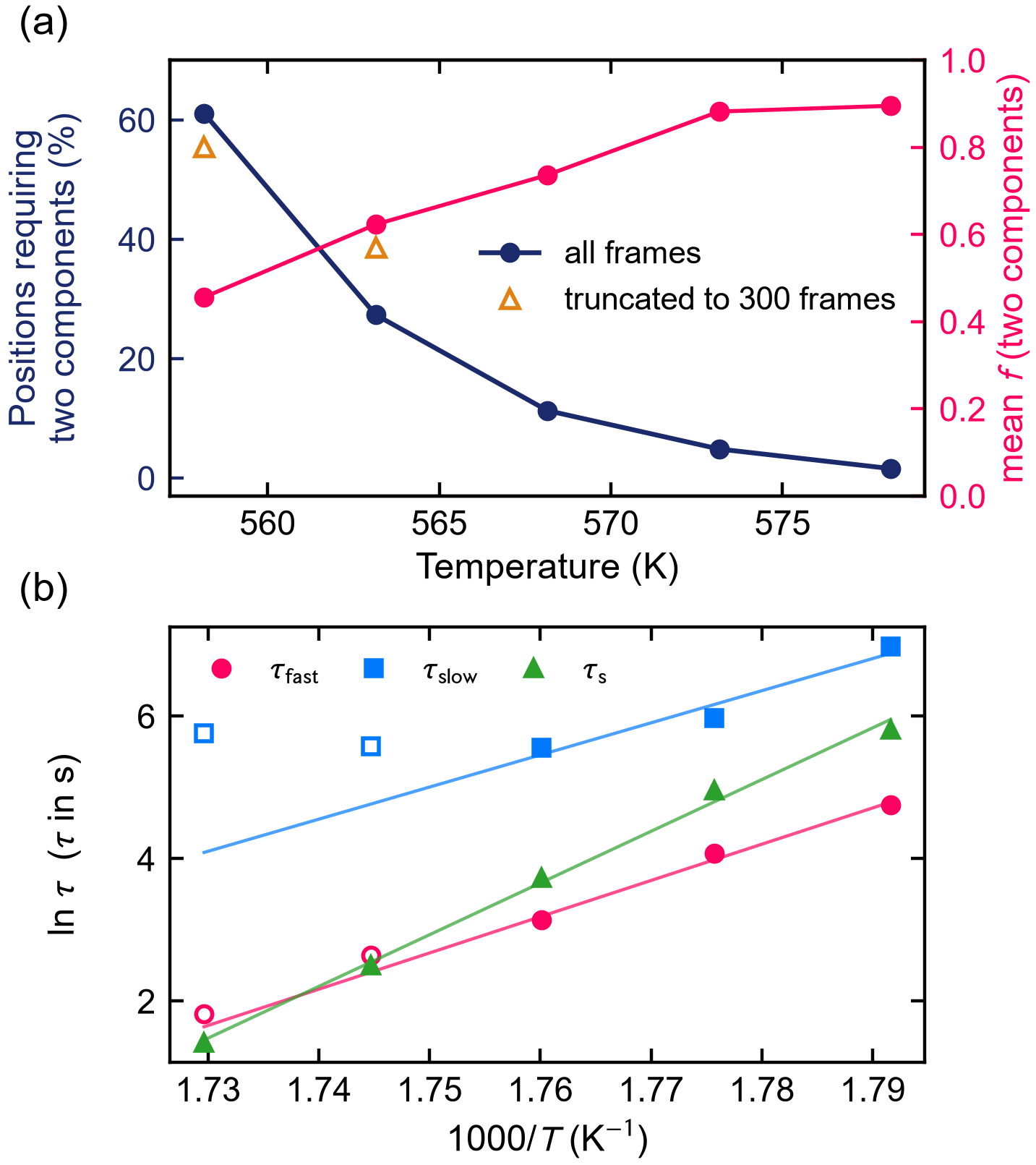}
\caption{(a) Fraction of positions requiring two components as a function of temperature (blue
circles, left axis). Orange triangles show the values obtained when the data at 558 and 563~K are
truncated to 300 frames, the number acquired at 568~K and above. Red circles (right axis) show the mean
$f$ at the two-component positions. (b) Arrhenius plot of the logarithmic averages of
$\taufast$ (red circles) and $\tauslow$ (blue squares) at the two-component positions and of $\taus$
(green triangles) at the one-component positions. Lines are fits over 558--568~K for $\taufast$ and
$\tauslow$ and over 558--578~K for $\taus$. Open symbols at 573~K and above are obtained from fewer
than 150 positions and are shown for reference only.}
\label{fig:temp}
\end{figure}

The fraction of two-component positions decreases monotonically with temperature: 61\%, 27\%, 11\%, 5\% and
2\% at 558, 563, 568, 573 and 578~K, respectively [Fig.~\ref{fig:temp}(a)]. At 578~K about 7\% of the
analyzed region, at its lower-left corner, has crystallized (Sec.~III~B of the Supplemental Material);
excluding these positions changes neither the fraction of two-component positions (1.6\% and 1.5\%) nor
$\taus$ (4.1 and 4.2~s) appreciably. Because fewer frames were
acquired above 568~K (300 frames instead of 600), we examined whether this decrease was caused by the
measurement conditions. When the data at 558 and 563~K were truncated to 300 frames, the fraction
remained 55--56\% and 38--39\%, respectively, far above the 11\% at 568~K obtained with the same number
of frames. The decrease was therefore caused by the temperature, not the acquisition time. At the positions that
still require two components, the mean $f$ increases from 0.46 at 558~K to 0.90 at 578~K
[Fig.~\ref{fig:temp}(a)], so the fast component also comes to dominate the relaxation where two
components remain. The two-component character of the relaxation is thus lost as the temperature
approaches $\Tg$ in two ways at once. This is consistent with the change from a one-component to a two-component decay upon cooling observed by XPCS in the glassy state of metallic glasses~\cite{Luo2017}.

Figure~\ref{fig:temp}(b) shows the temperature dependence of the logarithmic averages of $\taufast$ and
$\tauslow$ at the two-component positions and of $\taus$ at the one-component positions. Between 558 and 568~K,
$\taufast$ decreases from 116 to 23~s and $\tauslow$ from $1.1\times10^{3}$ to $2.6\times10^{2}$~s;
at 573~K and above, where only a few positions are classified as two-component, these values are shown for reference
only. $\taus$ decreases from 335~s at 558~K to 4.1~s at 578~K. The apparent activation energies are
$423 \pm 40$~kJ\,mol$^{-1}$ for $\taufast$ (558--568~K) and $604 \pm 23$~kJ\,mol$^{-1}$ for $\taus$
(558--578~K). The activation energy of $\tauslow$ ($375 \pm 87$~kJ\,mol$^{-1}$) is uncertain, because it
depends strongly on the selection of positions and temperatures; we therefore do not discuss the
difference between the temperature dependences of $\taufast$ and $\tauslow$.

We next consider the origin of the two components. For the present composition, mechanical spectroscopy
gives an activation energy of the slow $\beta$ (Johari--Goldstein) relaxation of $E_{\beta} \approx
120$--$130$~kJ\,mol$^{-1}$ ($E_{\beta}/R\Tg = 25$--28)~\cite{Qiao2014}, and the $\beta$-relaxation time
near $\Tg$ is in the $\mu$s--ms range~\cite{Qiao2014,Yu2013,Liu2016}, so that $\tau_{\beta} \lesssim
10^{-3}$~s at 558--578~K ($0.98 \le T/\Tg \le 1.01$). The observed relaxation times are more than three
orders of magnitude longer, and the activation energies three to five times larger. In contrast, the
apparent activation energy of the structural ($\alpha$) relaxation estimated from the fragility index,
$E_{\alpha} \approx m \ln(10) R \Tg \approx 640$~kJ\,mol$^{-1}$, agrees well with that of $\taus$, and
the observed time scales are comparable to the $\alpha$-relaxation time at $\Tg$ (about 100~s). Neither
component therefore corresponds to the $\beta$ relaxation; both belong to the time scale of the
structural relaxation. Mechanical spectroscopy has resolved two peaks in the $\alpha$ relaxation of a
metallic glass, interpreted as two glass transitions~\cite{Okumura1992}, and the two components observed
here may be related to such a splitting of the structural relaxation.

\subsection{Relation to local structure}

\begin{figure*}[t]
\includegraphics[width=\textwidth]{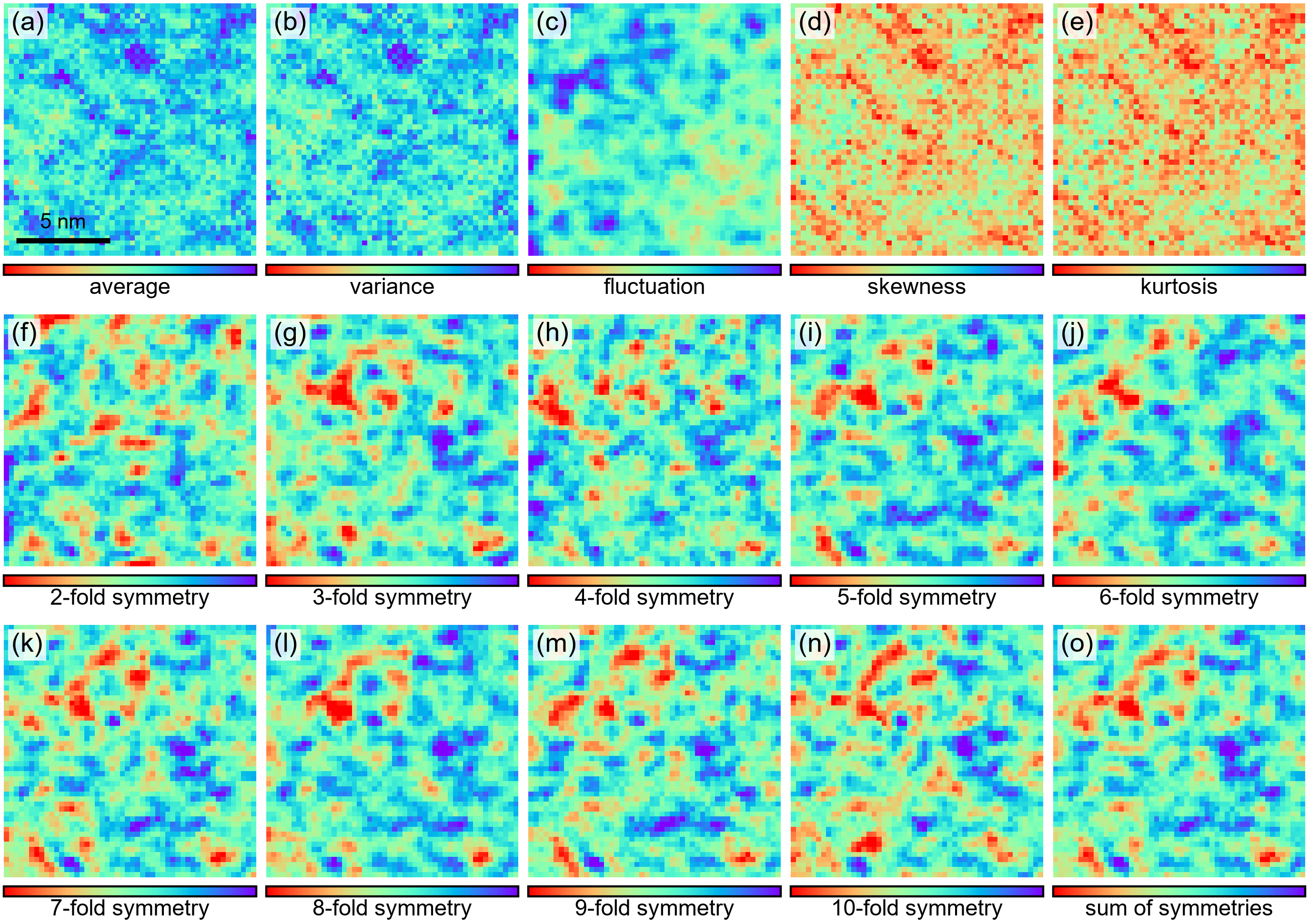}
\caption{Spatial distributions of the structural parameters at 558~K, averaged over the measurement:
(a) average, (b) variance, (c) fluctuation, (d) skewness, and (e) kurtosis of the diffracted intensity,
(f)--(n) 2- to 10-fold symmetries, and (o) the sum of the symmetries. The scale bar is 5~nm.}
\label{fig:struct}
\end{figure*}

We also evaluated the local atomic structure from the same diffraction patterns. As structural
parameters, we calculated the average, variance, fluctuation, skewness and kurtosis of the intensity,
and the $n$-fold symmetries. Fluctuation measures the intensity variation of the diffraction patterns
via
\begin{equation}
V(\bm{r},\bm{k})=
\frac{\left\langle I(\bm{r},\bm{k})^{2}\right\rangle_{\phi}}
     {\left\langle I(\bm{r},\bm{k})\right\rangle_{\phi}^{2}}-1
-\frac{1}{\left\langle I(\bm{r},\bm{k})\right\rangle_{\phi}} .
\label{eq:V}
\end{equation}
The last term removes the contribution of counting statistics to the normalized variance, so that $V$
measures the speckle contrast of the diffraction pattern alone. A higher value indicates
stronger speckles, and since regions with higher structural order tend to form stronger speckles, a
higher value indicates a more ordered structure~\cite{Hirata2014,Hirata2021}. We also assessed the symmetries in the diffraction patterns. The
$n$-fold symmetries, called symmetry order $S_n$, were measured from the angular autocorrelation $\rho$ of
the intensity [Eq.~\eqref{eq:rho}] and its average $S$ over the angles $2\pi m/n$, where $m$ and $n$ are
coprime ($m \perp n$) [Eq.~\eqref{eq:S}]~\cite{Huang2022}. Here $k_d$ and $k_{\theta}$ indicate the radius and azimuth in
the diffraction pattern and $\Delta$ indicates the correlation angle. The same angular region as that used
for the relaxation time measurement was employed, and the parameters were averaged over time to obtain
two-dimensional real-space distributions. The structural parameters are also spatially heterogeneous
(Fig.~\ref{fig:struct}).
\begin{widetext}
\begin{equation}
\rho(\bm{r},\Delta,t)=
\frac{\left\langle I(\bm{r},k_d,k_{\theta},t)\,I(\bm{r},k_d,k_{\theta}+\Delta,t)\right\rangle_{\phi}
      -\left\langle I(\bm{r},k_d,k_{\theta},t)\right\rangle_{\phi}^{2}}
     {\left\langle I(\bm{r},k_d,k_{\theta},t)^{2}\right\rangle_{\phi}
      -\left\langle I(\bm{r},k_d,k_{\theta},t)\right\rangle_{\phi}^{2}} ,
\label{eq:rho}
\end{equation}
\begin{equation}
S(\bm{r},n,t)=
\frac{\displaystyle\sum_{\substack{m=1\\ m \perp n}}^{n-1}\rho\!\left(\bm{r},\frac{2\pi m}{n},t\right)}
     {\displaystyle\sum_{\substack{m=1\\ m \perp n}}^{n-1} 1} .
\label{eq:S}
\end{equation}
\end{widetext}

The correlations between the structural and relaxation parameters were evaluated by Spearman's rank
correlation, with the significance assessed by a toroidal-shift surrogate test, which preserves the
spatial autocorrelation of each map, followed by a false-discovery-rate correction. Because all
structural parameters are themselves correlated with the fluctuation $V$, Fig.~\ref{fig:corr} shows the
partial rank correlations with $V$ controlled for (Sec.~V of the Supplemental Material). The most robust
correlation is the negative correlation between $f$ and $V$, which is significant at 558, 563 and 568~K
($\rho = -0.19$, $-0.12$ and $-0.11$, respectively) and persists when the analysis is restricted to the
two-component positions ($\rho = -0.21$ at 558~K). $V$ is also positively correlated with $\taufast$ at
558--573~K ($\rho = 0.09$--$0.16$), significantly after the false-discovery-rate correction at 563--573~K
($p = 0.02$ before the correction at 558~K). Once $V$ is controlled for, none of the other structural parameters,
including the $n$-fold symmetries, shows a significant correlation with any relaxation parameter at
558--568~K.
The rank correlations are small mainly because the position-wise $f$ is noisy and because the structure
changes during the measurement (Sec.~III of the Supplemental Material).
To express the dependence on the fluctuation in relaxation parameters rather than as a rank correlation, we divided the
positions at 558~K into five equal groups according to the fluctuation and compared the two-component
positions in the group with the lowest fluctuation (least ordered) with those in the group with the
highest fluctuation (most ordered). From the former to the latter, the median $f$ decreases from 0.55 to
0.38 and the median $\taufast$ increases from 87 to 135~s.

\begin{figure*}[t]
\includegraphics[width=\textwidth]{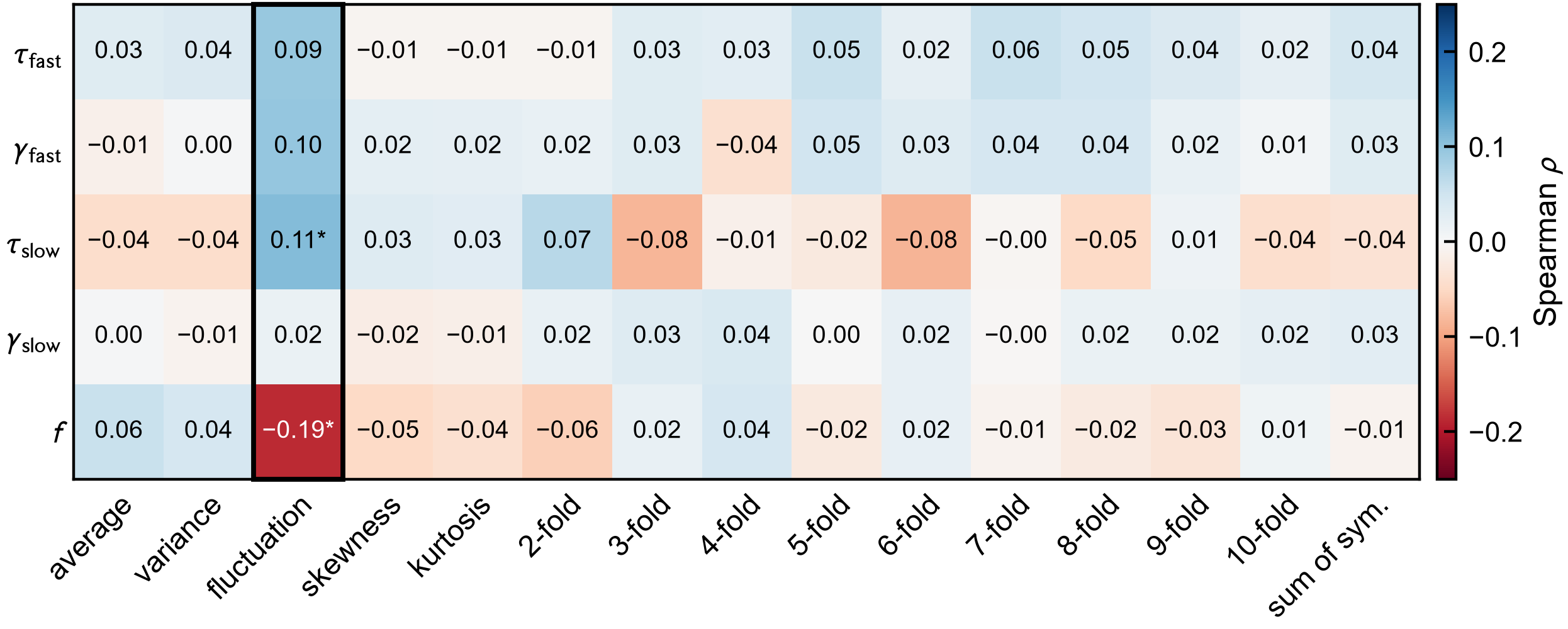}
\caption{Correlations between the relaxation and the structural parameters at 558~K with the fluctuation
$V$ controlled for. Each cell gives the partial Spearman rank correlation with $V$ controlled for, except
for the $V$ column itself (boxed), which gives the Spearman rank correlation. Blue indicates positive and
red negative correlation. Asterisks mark the correlations that remain significant after the
toroidal-shift surrogate test and the false-discovery-rate correction. The correlations before the
control are shown in Fig.~S21 in the Supplemental Material.}
\label{fig:corr}
\end{figure*}

\section{Discussion}

The present results show that the local relaxation is not described by one stretched decay: at
more than half of the positions at 558~K, a second component is statistically required even though the
one-component KWW function already allows for stretching. The local relaxation therefore consists of two
distinct processes. This cannot be recognized in spatially averaged measurements, in which the
superposition of many local decays appears as one stretched decay.

The scale at which this statement holds is set by the probed volume. The probe size is 0.8~nm in FWHM
and the scan step is 0.27~nm. If the two
components arose from laterally adjacent regions larger than the probe, each relaxing with one
component, two-component positions would appear only along the boundaries between these regions.
Instead, they form extended domains covering more than half of the analyzed area at 558~K
[Figs.~\ref{fig:maps}(a)--\ref{fig:maps}(c); Fig.~S6(i) in the Supplemental Material]. The two-component relaxation therefore occurs within volumes less than a nanometer
wide in the lateral direction. Along the beam direction, however, the signal is averaged over the sample thickness of 50--60~nm, so the
two components could in principle arise from regions with different dynamics stacked along this
direction. In that case the number of stacked regions, and hence the fraction of two-component positions
and the distribution of $f$, would vary systematically with thickness. This can be tested only at 533~K,
because at higher temperatures the thin part of the field of view has crystallized. At 533~K the
thickness varies from 15 to 65~nm within the field of view, and no such dependence is found (Sec.~I and
Fig.~S2 in the Supplemental Material), although the slow component at this temperature lies beyond the
observation time. This suggests that the two-component relaxation is not a result of stacking but is
local along the beam direction as well.

Positions with a higher degree of structural order show a larger contribution from the slow component and a
longer $\taufast$. This is consistent with the view that locally ordered structures are energetically
more stable and relax more slowly~\cite{Tong2018}, and with the picture of fragile metallic glasses as a
mixture of strongly and weakly bonded regions with different mobilities~\cite{Ichitsubo2005,Ichitsubo2011}.
One possible interpretation is that a less-ordered state relaxes faster and a more ordered
state relaxes more slowly, so that $f$ reflects the weight of the less-ordered state. The two states may
coexist within the probed volume, mixed on the nanometer scale, or the structure at each position may
alternate between them during the measurement; the latter is consistent with the change of the structure
during the measurement and with the finding that $f$ correlates with the fluctuation only within the same
time interval (Sec.~III of the Supplemental Material). Although the rank correlations are weak
($|\rho| \le 0.21$), they are likely to underestimate the underlying correlation.

The analysis also has implications for how local relaxation data are evaluated. A one-component KWW
function fitted to a local correlation function that contains two components returns an apparent
stretching exponent that mixes them. A position classified as one-component is one at which a second
component is not statistically required at the present noise level, not necessarily one at which it is
absent; the fraction of two-component positions at 558~K, for example, decreases from 61\% to 55--56\% when the data are
truncated to 300 frames. The strongly stretched decay at the one-component positions may therefore partly
reflect undetected two-component relaxation and should not be read as the stretching of one process
everywhere. The same caution applies to electron correlation microscopy and to local analyses of XPCS
data.

The main limitations of the present work are the averaging along the beam direction, the small number of
two-component positions at 573~K and above, which makes the time scales of the two components uncertain there, and the weakness of the
structure--dynamics correlations, although they remain significant after accounting for spatial
autocorrelation.

\section{Conclusion}

Using 5D-STEM, we determined position by position whether the structural relaxation of a fragile
metallic glass near $\Tg$ proceeds in one or two components. At 558~K, 61\% of the positions require two
components, so that the relaxation within nanometer-scale volumes consists of two distinct processes
rather than one stretched decay. Both components lie on the time scale of the structural relaxation
rather than of the $\beta$ relaxation. As the temperature is increased, the fraction of two-component positions
decreases to 2\% at 578~K, and the local relaxation changes from two-component to one-component.
Positions with higher structural order show a larger contribution from the slow component and a longer
$\taufast$. These correlations are weak, but they are likely to underestimate the underlying
correlation, because the structure changes during the measurement. The two-component relaxation is
thus linked to the local structure.

\begin{acknowledgments}
This work was supported by ``Advanced Research Infrastructure for Materials and Nanotechnology in Japan
(ARIM)'' of the Ministry of Education, Culture, Sports, Science and Technology (MEXT) proposal number
JPMXP1225NM5126 and JSPS KAKENHI grant number 23K13076.
\end{acknowledgments}

\bibliography{refs}

\begin{thebibliography}{34}%
\makeatletter
\providecommand \@ifxundefined [1]{%
 \@ifx{#1\undefined}
}%
\providecommand \@ifnum [1]{%
 \ifnum #1\expandafter \@firstoftwo
 \else \expandafter \@secondoftwo
 \fi
}%
\providecommand \@ifx [1]{%
 \ifx #1\expandafter \@firstoftwo
 \else \expandafter \@secondoftwo
 \fi
}%
\providecommand \natexlab [1]{#1}%
\providecommand \enquote  [1]{``#1''}%
\providecommand \bibnamefont  [1]{#1}%
\providecommand \bibfnamefont [1]{#1}%
\providecommand \citenamefont [1]{#1}%
\providecommand \href@noop [0]{\@secondoftwo}%
\providecommand \href [0]{\begingroup \@sanitize@url \@href}%
\providecommand \@href[1]{\@@startlink{#1}\@@href}%
\providecommand \@@href[1]{\endgroup#1\@@endlink}%
\providecommand \@sanitize@url [0]{\catcode `\\12\catcode `\$12\catcode
  `\&12\catcode `\#12\catcode `\^12\catcode `\_12\catcode `\%12\relax}%
\providecommand \@@startlink[1]{}%
\providecommand \@@endlink[0]{}%
\providecommand \url  [0]{\begingroup\@sanitize@url \@url }%
\providecommand \@url [1]{\endgroup\@href {#1}{\urlprefix }}%
\providecommand \urlprefix  [0]{URL }%
\providecommand \Eprint [0]{\href }%
\providecommand \doibase [0]{https://doi.org/}%
\providecommand \selectlanguage [0]{\@gobble}%
\providecommand \bibinfo  [0]{\@secondoftwo}%
\providecommand \bibfield  [0]{\@secondoftwo}%
\providecommand \translation [1]{[#1]}%
\providecommand \BibitemOpen [0]{}%
\providecommand \bibitemStop [0]{}%
\providecommand \bibitemNoStop [0]{.\EOS\space}%
\providecommand \EOS [0]{\spacefactor3000\relax}%
\providecommand \BibitemShut  [1]{\csname bibitem#1\endcsname}%
\let\auto@bib@innerbib\@empty
\bibitem [{\citenamefont {Ediger}\ \emph {et~al.}(1996)\citenamefont {Ediger},
  \citenamefont {Angell},\ and\ \citenamefont {Nagel}}]{Ediger1996}%
  \BibitemOpen
  \bibfield  {author} {\bibinfo {author} {\bibfnamefont {M.~D.}\ \bibnamefont
  {Ediger}}, \bibinfo {author} {\bibfnamefont {C.~A.}\ \bibnamefont {Angell}},\
  and\ \bibinfo {author} {\bibfnamefont {S.~R.}\ \bibnamefont {Nagel}},\
  }\bibfield  {title} {\bibinfo {title} {{Supercooled Liquids and Glasses}},\
  }\href {https://doi.org/10.1021/jp953538d} {\bibfield  {journal} {\bibinfo
  {journal} {J. Phys. Chem.}\ }\textbf {\bibinfo {volume} {100}},\ \bibinfo
  {pages} {13200} (\bibinfo {year} {1996})}\BibitemShut {NoStop}%
\bibitem [{\citenamefont {Berthier}\ and\ \citenamefont
  {Biroli}(2011)}]{Berthier2011}%
  \BibitemOpen
  \bibfield  {author} {\bibinfo {author} {\bibfnamefont {L.}~\bibnamefont
  {Berthier}}\ and\ \bibinfo {author} {\bibfnamefont {G.}~\bibnamefont
  {Biroli}},\ }\bibfield  {title} {\bibinfo {title} {{Theoretical perspective
  on the glass transition and amorphous materials}},\ }\href
  {https://doi.org/10.1103/RevModPhys.83.587} {\bibfield  {journal} {\bibinfo
  {journal} {Rev. Mod. Phys.}\ }\textbf {\bibinfo {volume} {83}},\ \bibinfo
  {pages} {587} (\bibinfo {year} {2011})}\BibitemShut {NoStop}%
\bibitem [{\citenamefont {Angell}(1988)}]{Angell1988}%
  \BibitemOpen
  \bibfield  {author} {\bibinfo {author} {\bibfnamefont {C.~A.}\ \bibnamefont
  {Angell}},\ }\bibfield  {title} {\bibinfo {title} {{Perspective on the glass
  transition}},\ }\href {https://doi.org/10.1016/0022-3697(88)90002-9}
  {\bibfield  {journal} {\bibinfo  {journal} {J. Phys. Chem. Solids}\ }\textbf
  {\bibinfo {volume} {49}},\ \bibinfo {pages} {863} (\bibinfo {year}
  {1988})}\BibitemShut {NoStop}%
\bibitem [{\citenamefont {Angell}(1995)}]{Angell1995}%
  \BibitemOpen
  \bibfield  {author} {\bibinfo {author} {\bibfnamefont {C.~A.}\ \bibnamefont
  {Angell}},\ }\bibfield  {title} {\bibinfo {title} {{Formation of Glasses from
  Liquids and Biopolymers}},\ }\href
  {https://doi.org/10.1126/science.267.5206.1924} {\bibfield  {journal}
  {\bibinfo  {journal} {Science}\ }\textbf {\bibinfo {volume} {267}},\ \bibinfo
  {pages} {1924} (\bibinfo {year} {1995})}\BibitemShut {NoStop}%
\bibitem [{\citenamefont {Debenedetti}\ and\ \citenamefont
  {Stillinger}(2001)}]{Debenedetti2001}%
  \BibitemOpen
  \bibfield  {author} {\bibinfo {author} {\bibfnamefont {P.~G.}\ \bibnamefont
  {Debenedetti}}\ and\ \bibinfo {author} {\bibfnamefont {F.~H.}\ \bibnamefont
  {Stillinger}},\ }\bibfield  {title} {\bibinfo {title} {{Supercooled liquids
  and the glass transition}},\ }\href {https://doi.org/10.1038/35065704}
  {\bibfield  {journal} {\bibinfo  {journal} {Nature}\ }\textbf {\bibinfo
  {volume} {410}},\ \bibinfo {pages} {259} (\bibinfo {year}
  {2001})}\BibitemShut {NoStop}%
\bibitem [{\citenamefont {B{\"o}hmer}\ \emph {et~al.}(1993)\citenamefont
  {B{\"o}hmer}, \citenamefont {Ngai}, \citenamefont {Angell},\ and\
  \citenamefont {Plazek}}]{Bohmer1993}%
  \BibitemOpen
  \bibfield  {author} {\bibinfo {author} {\bibfnamefont {R.}~\bibnamefont
  {B{\"o}hmer}}, \bibinfo {author} {\bibfnamefont {K.~L.}\ \bibnamefont
  {Ngai}}, \bibinfo {author} {\bibfnamefont {C.~A.}\ \bibnamefont {Angell}},\
  and\ \bibinfo {author} {\bibfnamefont {D.~J.}\ \bibnamefont {Plazek}},\
  }\bibfield  {title} {\bibinfo {title} {{Nonexponential relaxations in strong
  and fragile glass formers}},\ }\href {https://doi.org/10.1063/1.466117}
  {\bibfield  {journal} {\bibinfo  {journal} {J. Chem. Phys.}\ }\textbf
  {\bibinfo {volume} {99}},\ \bibinfo {pages} {4201} (\bibinfo {year}
  {1993})}\BibitemShut {NoStop}%
\bibitem [{\citenamefont {Ediger}(2000)}]{Ediger2000}%
  \BibitemOpen
  \bibfield  {author} {\bibinfo {author} {\bibfnamefont {M.~D.}\ \bibnamefont
  {Ediger}},\ }\bibfield  {title} {\bibinfo {title} {{Spatially heterogeneous
  dynamics in supercooled liquids}},\ }\href
  {https://doi.org/10.1146/annurev.physchem.51.1.99} {\bibfield  {journal}
  {\bibinfo  {journal} {Annu. Rev. Phys. Chem.}\ }\textbf {\bibinfo {volume}
  {51}},\ \bibinfo {pages} {99} (\bibinfo {year} {2000})}\BibitemShut {NoStop}%
\bibitem [{\citenamefont {Richert}(2002)}]{Richert2002}%
  \BibitemOpen
  \bibfield  {author} {\bibinfo {author} {\bibfnamefont {R.}~\bibnamefont
  {Richert}},\ }\bibfield  {title} {\bibinfo {title} {{Heterogeneous dynamics
  in liquids: fluctuations in space and time}},\ }\href
  {https://doi.org/10.1088/0953-8984/14/23/201} {\bibfield  {journal} {\bibinfo
   {journal} {J. Phys.: Condens. Matter}\ }\textbf {\bibinfo {volume} {14}},\
  \bibinfo {pages} {R703} (\bibinfo {year} {2002})}\BibitemShut {NoStop}%
\bibitem [{\citenamefont {Luo}\ \emph {et~al.}(2017)\citenamefont {Luo},
  \citenamefont {Wen}, \citenamefont {Bai}, \citenamefont {Ruta},\ and\
  \citenamefont {Wang}}]{Luo2017}%
  \BibitemOpen
  \bibfield  {author} {\bibinfo {author} {\bibfnamefont {P.}~\bibnamefont
  {Luo}}, \bibinfo {author} {\bibfnamefont {P.}~\bibnamefont {Wen}}, \bibinfo
  {author} {\bibfnamefont {H.~Y.}\ \bibnamefont {Bai}}, \bibinfo {author}
  {\bibfnamefont {B.}~\bibnamefont {Ruta}},\ and\ \bibinfo {author}
  {\bibfnamefont {W.~H.}\ \bibnamefont {Wang}},\ }\bibfield  {title} {\bibinfo
  {title} {{Relaxation Decoupling in Metallic Glasses at Low Temperatures}},\
  }\href {https://doi.org/10.1103/PhysRevLett.118.225901} {\bibfield  {journal}
  {\bibinfo  {journal} {Phys. Rev. Lett.}\ }\textbf {\bibinfo {volume} {118}},\
  \bibinfo {pages} {225901} (\bibinfo {year} {2017})}\BibitemShut {NoStop}%
\bibitem [{\citenamefont {Ruta}\ \emph {et~al.}(2012)\citenamefont {Ruta},
  \citenamefont {Chushkin}, \citenamefont {Monaco}, \citenamefont {Cipelletti},
  \citenamefont {Pineda}, \citenamefont {Bruna}, \citenamefont {Giordano},\
  and\ \citenamefont {Gonzalez-Silveira}}]{Ruta2012}%
  \BibitemOpen
  \bibfield  {author} {\bibinfo {author} {\bibfnamefont {B.}~\bibnamefont
  {Ruta}}, \bibinfo {author} {\bibfnamefont {Y.}~\bibnamefont {Chushkin}},
  \bibinfo {author} {\bibfnamefont {G.}~\bibnamefont {Monaco}}, \bibinfo
  {author} {\bibfnamefont {L.}~\bibnamefont {Cipelletti}}, \bibinfo {author}
  {\bibfnamefont {E.}~\bibnamefont {Pineda}}, \bibinfo {author} {\bibfnamefont
  {P.}~\bibnamefont {Bruna}}, \bibinfo {author} {\bibfnamefont {V.~M.}\
  \bibnamefont {Giordano}},\ and\ \bibinfo {author} {\bibfnamefont
  {M.}~\bibnamefont {Gonzalez-Silveira}},\ }\bibfield  {title} {\bibinfo
  {title} {{Atomic-scale relaxation dynamics and aging in a metallic glass
  probed by X-ray photon correlation spectroscopy}},\ }\href
  {https://doi.org/10.1103/PhysRevLett.109.165701} {\bibfield  {journal}
  {\bibinfo  {journal} {Phys. Rev. Lett.}\ }\textbf {\bibinfo {volume} {109}},\
  \bibinfo {pages} {165701} (\bibinfo {year} {2012})}\BibitemShut {NoStop}%
\bibitem [{\citenamefont {Evenson}\ \emph {et~al.}(2015)\citenamefont
  {Evenson}, \citenamefont {Ruta}, \citenamefont {Hechler}, \citenamefont
  {Stolpe}, \citenamefont {Pineda}, \citenamefont {Gallino},\ and\
  \citenamefont {Busch}}]{Evenson2015}%
  \BibitemOpen
  \bibfield  {author} {\bibinfo {author} {\bibfnamefont {Z.}~\bibnamefont
  {Evenson}}, \bibinfo {author} {\bibfnamefont {B.}~\bibnamefont {Ruta}},
  \bibinfo {author} {\bibfnamefont {S.}~\bibnamefont {Hechler}}, \bibinfo
  {author} {\bibfnamefont {M.}~\bibnamefont {Stolpe}}, \bibinfo {author}
  {\bibfnamefont {E.}~\bibnamefont {Pineda}}, \bibinfo {author} {\bibfnamefont
  {I.}~\bibnamefont {Gallino}},\ and\ \bibinfo {author} {\bibfnamefont
  {R.}~\bibnamefont {Busch}},\ }\bibfield  {title} {\bibinfo {title} {{X-ray
  photon correlation spectroscopy reveals intermittent aging dynamics in a
  metallic glass}},\ }\href {https://doi.org/10.1103/PhysRevLett.115.175701}
  {\bibfield  {journal} {\bibinfo  {journal} {Phys. Rev. Lett.}\ }\textbf
  {\bibinfo {volume} {115}},\ \bibinfo {pages} {175701} (\bibinfo {year}
  {2015})}\BibitemShut {NoStop}%
\bibitem [{\citenamefont {Okumura}\ \emph {et~al.}(1992)\citenamefont
  {Okumura}, \citenamefont {Chen}, \citenamefont {Inoue},\ and\ \citenamefont
  {Masumoto}}]{Okumura1992}%
  \BibitemOpen
  \bibfield  {author} {\bibinfo {author} {\bibfnamefont {H.}~\bibnamefont
  {Okumura}}, \bibinfo {author} {\bibfnamefont {H.~S.}\ \bibnamefont {Chen}},
  \bibinfo {author} {\bibfnamefont {A.}~\bibnamefont {Inoue}},\ and\ \bibinfo
  {author} {\bibfnamefont {T.}~\bibnamefont {Masumoto}},\ }\bibfield  {title}
  {\bibinfo {title} {{Two glass transitions in mechanical relaxation of a
  La$_{55}$Al$_{25}$Ni$_{20}$ amorphous alloy}},\ }\href
  {https://doi.org/10.1016/S0022-3093(05)80020-1} {\bibfield  {journal}
  {\bibinfo  {journal} {J. Non-Cryst. Solids}\ }\textbf {\bibinfo {volume}
  {142}},\ \bibinfo {pages} {165} (\bibinfo {year} {1992})}\BibitemShut
  {NoStop}%
\bibitem [{\citenamefont {Tong}\ and\ \citenamefont {Tanaka}(2018)}]{Tong2018}%
  \BibitemOpen
  \bibfield  {author} {\bibinfo {author} {\bibfnamefont {H.}~\bibnamefont
  {Tong}}\ and\ \bibinfo {author} {\bibfnamefont {H.}~\bibnamefont {Tanaka}},\
  }\bibfield  {title} {\bibinfo {title} {{Revealing Hidden Structural Order
  Controlling Both Fast and Slow Glassy Dynamics in Supercooled Liquids}},\
  }\href {https://doi.org/10.1103/PhysRevX.8.011041} {\bibfield  {journal}
  {\bibinfo  {journal} {Phys. Rev. X}\ }\textbf {\bibinfo {volume} {8}},\
  \bibinfo {pages} {011041} (\bibinfo {year} {2018})}\BibitemShut {NoStop}%
\bibitem [{\citenamefont {Ichitsubo}\ \emph {et~al.}(2005)\citenamefont
  {Ichitsubo}, \citenamefont {Matsubara}, \citenamefont {Yamamoto},
  \citenamefont {Chen}, \citenamefont {Nishiyama}, \citenamefont {Saida},\ and\
  \citenamefont {Anazawa}}]{Ichitsubo2005}%
  \BibitemOpen
  \bibfield  {author} {\bibinfo {author} {\bibfnamefont {T.}~\bibnamefont
  {Ichitsubo}}, \bibinfo {author} {\bibfnamefont {E.}~\bibnamefont
  {Matsubara}}, \bibinfo {author} {\bibfnamefont {T.}~\bibnamefont {Yamamoto}},
  \bibinfo {author} {\bibfnamefont {H.~S.}\ \bibnamefont {Chen}}, \bibinfo
  {author} {\bibfnamefont {N.}~\bibnamefont {Nishiyama}}, \bibinfo {author}
  {\bibfnamefont {J.}~\bibnamefont {Saida}},\ and\ \bibinfo {author}
  {\bibfnamefont {K.}~\bibnamefont {Anazawa}},\ }\bibfield  {title} {\bibinfo
  {title} {{Microstructure of fragile metallic glasses inferred from
  ultrasound-accelerated crystallization in Pd-based metallic glasses}},\
  }\href {https://doi.org/10.1103/PhysRevLett.95.245501} {\bibfield  {journal}
  {\bibinfo  {journal} {Phys. Rev. Lett.}\ }\textbf {\bibinfo {volume} {95}},\
  \bibinfo {pages} {245501} (\bibinfo {year} {2005})}\BibitemShut {NoStop}%
\bibitem [{\citenamefont {Ichitsubo}\ \emph {et~al.}(2011)\citenamefont
  {Ichitsubo}, \citenamefont {Kato}, \citenamefont {Matsubara}, \citenamefont
  {Biwa}, \citenamefont {Hosokawa}, \citenamefont {Matsuda}, \citenamefont
  {Uchiyama},\ and\ \citenamefont {Baron}}]{Ichitsubo2011}%
  \BibitemOpen
  \bibfield  {author} {\bibinfo {author} {\bibfnamefont {T.}~\bibnamefont
  {Ichitsubo}}, \bibinfo {author} {\bibfnamefont {H.}~\bibnamefont {Kato}},
  \bibinfo {author} {\bibfnamefont {E.}~\bibnamefont {Matsubara}}, \bibinfo
  {author} {\bibfnamefont {S.}~\bibnamefont {Biwa}}, \bibinfo {author}
  {\bibfnamefont {S.}~\bibnamefont {Hosokawa}}, \bibinfo {author}
  {\bibfnamefont {K.}~\bibnamefont {Matsuda}}, \bibinfo {author} {\bibfnamefont
  {H.}~\bibnamefont {Uchiyama}},\ and\ \bibinfo {author} {\bibfnamefont
  {A.~Q.~R.}\ \bibnamefont {Baron}},\ }\bibfield  {title} {\bibinfo {title}
  {{Static heterogeneity in metallic glasses and its correlation to physical
  properties}},\ }\href {https://doi.org/10.1016/j.jnoncrysol.2010.06.056}
  {\bibfield  {journal} {\bibinfo  {journal} {J. Non-Cryst. Solids}\ }\textbf
  {\bibinfo {volume} {357}},\ \bibinfo {pages} {494} (\bibinfo {year}
  {2011})}\BibitemShut {NoStop}%
\bibitem [{\citenamefont {Spangenberg}\ \emph {et~al.}(2021)\citenamefont
  {Spangenberg}, \citenamefont {Hilke}, \citenamefont {Wilde},\ and\
  \citenamefont {Peterlechner}}]{Spangenberg2021}%
  \BibitemOpen
  \bibfield  {author} {\bibinfo {author} {\bibfnamefont {K.}~\bibnamefont
  {Spangenberg}}, \bibinfo {author} {\bibfnamefont {S.}~\bibnamefont {Hilke}},
  \bibinfo {author} {\bibfnamefont {G.}~\bibnamefont {Wilde}},\ and\ \bibinfo
  {author} {\bibfnamefont {M.}~\bibnamefont {Peterlechner}},\ }\bibfield
  {title} {\bibinfo {title} {{Direct View on Non-Equilibrium Heterogeneous
  Dynamics in Glassy Nanorods}},\ }\href
  {https://doi.org/10.1002/adfm.202103742} {\bibfield  {journal} {\bibinfo
  {journal} {Adv. Funct. Mater.}\ }\textbf {\bibinfo {volume} {31}},\ \bibinfo
  {pages} {2103742} (\bibinfo {year} {2021})}\BibitemShut {NoStop}%
\bibitem [{\citenamefont {Zhang}\ \emph {et~al.}(2018)\citenamefont {Zhang},
  \citenamefont {Maldonis}, \citenamefont {Liu}, \citenamefont {Schroers},\
  and\ \citenamefont {Voyles}}]{Zhang2018}%
  \BibitemOpen
  \bibfield  {author} {\bibinfo {author} {\bibfnamefont {P.}~\bibnamefont
  {Zhang}}, \bibinfo {author} {\bibfnamefont {J.~J.}\ \bibnamefont {Maldonis}},
  \bibinfo {author} {\bibfnamefont {Z.}~\bibnamefont {Liu}}, \bibinfo {author}
  {\bibfnamefont {J.}~\bibnamefont {Schroers}},\ and\ \bibinfo {author}
  {\bibfnamefont {P.~M.}\ \bibnamefont {Voyles}},\ }\bibfield  {title}
  {\bibinfo {title} {{Spatially heterogeneous dynamics in a metallic glass
  forming liquid imaged by electron correlation microscopy}},\ }\href
  {https://doi.org/10.1038/s41467-018-03604-2} {\bibfield  {journal} {\bibinfo
  {journal} {Nat. Commun.}\ }\textbf {\bibinfo {volume} {9}},\ \bibinfo {pages}
  {1129} (\bibinfo {year} {2018})}\BibitemShut {NoStop}%
\bibitem [{\citenamefont {Nakazawa}\ \emph {et~al.}(2024)\citenamefont
  {Nakazawa}, \citenamefont {Mitsuishi}, \citenamefont {Iakoubovskii},
  \citenamefont {Kohara},\ and\ \citenamefont {Tsuchiya}}]{Nakazawa2024}%
  \BibitemOpen
  \bibfield  {author} {\bibinfo {author} {\bibfnamefont {K.}~\bibnamefont
  {Nakazawa}}, \bibinfo {author} {\bibfnamefont {K.}~\bibnamefont {Mitsuishi}},
  \bibinfo {author} {\bibfnamefont {K.}~\bibnamefont {Iakoubovskii}}, \bibinfo
  {author} {\bibfnamefont {S.}~\bibnamefont {Kohara}},\ and\ \bibinfo {author}
  {\bibfnamefont {K.}~\bibnamefont {Tsuchiya}},\ }\bibfield  {title} {\bibinfo
  {title} {{Structure-dynamics relation in metallic glass revealed by
  5-dimensional scanning transmission electron microscopy}},\ }\href
  {https://doi.org/10.1038/s41427-024-00577-1} {\bibfield  {journal} {\bibinfo
  {journal} {NPG Asia Mater.}\ }\textbf {\bibinfo {volume} {16}},\ \bibinfo
  {pages} {57} (\bibinfo {year} {2024})}\BibitemShut {NoStop}%
\bibitem [{\citenamefont {Nakazawa}\ and\ \citenamefont
  {Mitsuishi}(2023)}]{Nakazawa2023}%
  \BibitemOpen
  \bibfield  {author} {\bibinfo {author} {\bibfnamefont {K.}~\bibnamefont
  {Nakazawa}}\ and\ \bibinfo {author} {\bibfnamefont {K.}~\bibnamefont
  {Mitsuishi}},\ }\bibfield  {title} {\bibinfo {title} {{Development of
  temporal series 4D-STEM and application to relaxation time measurement}},\
  }\href {https://doi.org/10.1093/jmicro/dfad006} {\bibfield  {journal}
  {\bibinfo  {journal} {Microscopy}\ }\textbf {\bibinfo {volume} {72}},\
  \bibinfo {pages} {446} (\bibinfo {year} {2023})}\BibitemShut {NoStop}%
\bibitem [{\citenamefont {Huang}\ and\ \citenamefont
  {Voyles}(2024)}]{Huang2024}%
  \BibitemOpen
  \bibfield  {author} {\bibinfo {author} {\bibfnamefont {S.}~\bibnamefont
  {Huang}}\ and\ \bibinfo {author} {\bibfnamefont {P.~M.}\ \bibnamefont
  {Voyles}},\ }\bibfield  {title} {\bibinfo {title} {{Momentum transfer
  resolved electron correlation microscopy}},\ }\href
  {https://doi.org/10.1016/j.ultramic.2023.113886} {\bibfield  {journal}
  {\bibinfo  {journal} {Ultramicroscopy}\ }\textbf {\bibinfo {volume} {256}},\
  \bibinfo {pages} {113886} (\bibinfo {year} {2024})}\BibitemShut {NoStop}%
\bibitem [{\citenamefont {Kato}\ \emph {et~al.}(2006)\citenamefont {Kato},
  \citenamefont {Wada}, \citenamefont {Hasegawa}, \citenamefont {Saida},
  \citenamefont {Inoue},\ and\ \citenamefont {Chen}}]{Kato2006}%
  \BibitemOpen
  \bibfield  {author} {\bibinfo {author} {\bibfnamefont {H.}~\bibnamefont
  {Kato}}, \bibinfo {author} {\bibfnamefont {T.}~\bibnamefont {Wada}}, \bibinfo
  {author} {\bibfnamefont {M.}~\bibnamefont {Hasegawa}}, \bibinfo {author}
  {\bibfnamefont {J.}~\bibnamefont {Saida}}, \bibinfo {author} {\bibfnamefont
  {A.}~\bibnamefont {Inoue}},\ and\ \bibinfo {author} {\bibfnamefont {H.~S.}\
  \bibnamefont {Chen}},\ }\bibfield  {title} {\bibinfo {title} {{Fragility and
  thermal stability of Pt- and Pd-based bulk glass forming liquids and their
  correlation with deformability}},\ }\href
  {https://doi.org/10.1016/j.scriptamat.2006.03.025} {\bibfield  {journal}
  {\bibinfo  {journal} {Scr. Mater.}\ }\textbf {\bibinfo {volume} {54}},\
  \bibinfo {pages} {2023} (\bibinfo {year} {2006})}\BibitemShut {NoStop}%
\bibitem [{\citenamefont {Haruyama}\ \emph {et~al.}(2011)\citenamefont
  {Haruyama}, \citenamefont {Watanabe}, \citenamefont {Yuki}, \citenamefont
  {Horiuchi}, \citenamefont {Kato},\ and\ \citenamefont
  {Nishiyama}}]{Haruyama2011}%
  \BibitemOpen
  \bibfield  {author} {\bibinfo {author} {\bibfnamefont {O.}~\bibnamefont
  {Haruyama}}, \bibinfo {author} {\bibfnamefont {T.}~\bibnamefont {Watanabe}},
  \bibinfo {author} {\bibfnamefont {K.}~\bibnamefont {Yuki}}, \bibinfo {author}
  {\bibfnamefont {M.}~\bibnamefont {Horiuchi}}, \bibinfo {author}
  {\bibfnamefont {H.}~\bibnamefont {Kato}},\ and\ \bibinfo {author}
  {\bibfnamefont {N.}~\bibnamefont {Nishiyama}},\ }\bibfield  {title} {\bibinfo
  {title} {{Thermodynamic approach to glass-forming ability of water-quenched
  Pd-P-based and Pt$_{60}$Ni$_{15}$P$_{25}$ bulk metallic glasses}},\ }\href
  {https://doi.org/10.1103/PhysRevB.83.064201} {\bibfield  {journal} {\bibinfo
  {journal} {Phys. Rev. B}\ }\textbf {\bibinfo {volume} {83}},\ \bibinfo
  {pages} {064201} (\bibinfo {year} {2011})}\BibitemShut {NoStop}%
\bibitem [{\citenamefont {Madsen}\ \emph {et~al.}(2010)\citenamefont {Madsen},
  \citenamefont {Leheny}, \citenamefont {Guo}, \citenamefont {Sprung},\ and\
  \citenamefont {Czakkel}}]{Madsen2010}%
  \BibitemOpen
  \bibfield  {author} {\bibinfo {author} {\bibfnamefont {A.}~\bibnamefont
  {Madsen}}, \bibinfo {author} {\bibfnamefont {R.~L.}\ \bibnamefont {Leheny}},
  \bibinfo {author} {\bibfnamefont {H.}~\bibnamefont {Guo}}, \bibinfo {author}
  {\bibfnamefont {M.}~\bibnamefont {Sprung}},\ and\ \bibinfo {author}
  {\bibfnamefont {O.}~\bibnamefont {Czakkel}},\ }\bibfield  {title} {\bibinfo
  {title} {{Beyond simple exponential correlation functions and equilibrium
  dynamics in x-ray photon correlation spectroscopy}},\ }\href
  {https://doi.org/10.1088/1367-2630/12/5/055001} {\bibfield  {journal}
  {\bibinfo  {journal} {New J. Phys.}\ }\textbf {\bibinfo {volume} {12}},\
  \bibinfo {pages} {055001} (\bibinfo {year} {2010})}\BibitemShut {NoStop}%
\bibitem [{\citenamefont {Nakazawa}\ and\ \citenamefont
  {Mitsuishi}(2025)}]{Nakazawa2025}%
  \BibitemOpen
  \bibfield  {author} {\bibinfo {author} {\bibfnamefont {K.}~\bibnamefont
  {Nakazawa}}\ and\ \bibinfo {author} {\bibfnamefont {K.}~\bibnamefont
  {Mitsuishi}},\ }\bibfield  {title} {\bibinfo {title} {{Relaxation time
  measurement: correlating diffraction patterns}},\ }\href
  {https://doi.org/10.1093/jmicro/dfaf048} {\bibfield  {journal} {\bibinfo
  {journal} {Microscopy}\ }\textbf {\bibinfo {volume} {75}},\ \bibinfo {pages}
  {285} (\bibinfo {year} {2025})}\BibitemShut {NoStop}%
\bibitem [{\citenamefont {Bikondoa}(2017)}]{Bikondoa2017}%
  \BibitemOpen
  \bibfield  {author} {\bibinfo {author} {\bibfnamefont {O.}~\bibnamefont
  {Bikondoa}},\ }\bibfield  {title} {\bibinfo {title} {{On the use of two-time
  correlation functions for X-ray photon correlation spectroscopy data
  analysis}},\ }\href {https://doi.org/10.1107/S1600576717000577} {\bibfield
  {journal} {\bibinfo  {journal} {J. Appl. Crystallogr.}\ }\textbf {\bibinfo
  {volume} {50}},\ \bibinfo {pages} {357} (\bibinfo {year} {2017})}\BibitemShut
  {NoStop}%
\bibitem [{SM()}]{SM}%
  \BibitemOpen
  \href@noop {} {}\bibinfo {note} {See the Supplemental Material appended to
  this article for the sample thickness, details of the model selection, the
  reasons for the weak structural correlations, and maps at all temperatures,
  which includes Refs.~\cite{Egerton2011,Chatterjee2021}}\BibitemShut {NoStop}%
\bibitem [{\citenamefont {Egerton}(2011)}]{Egerton2011}%
  \BibitemOpen
  \bibfield  {author} {\bibinfo {author} {\bibfnamefont {R.~F.}\ \bibnamefont
  {Egerton}},\ }\href {https://doi.org/10.1007/978-1-4419-9583-4} {\emph
  {\bibinfo {title} {{Electron Energy-Loss Spectroscopy in the Electron
  Microscope}}}},\ \bibinfo {edition} {3rd}\ ed.\ (\bibinfo  {publisher}
  {Springer},\ \bibinfo {address} {New York},\ \bibinfo {year}
  {2011})\BibitemShut {NoStop}%
\bibitem [{\citenamefont {Chatterjee}\ \emph {et~al.}(2021)\citenamefont
  {Chatterjee}, \citenamefont {Annamareddy}, \citenamefont {Ketkaew},
  \citenamefont {Schroers}, \citenamefont {Morgan},\ and\ \citenamefont
  {Voyles}}]{Chatterjee2021}%
  \BibitemOpen
  \bibfield  {author} {\bibinfo {author} {\bibfnamefont {D.}~\bibnamefont
  {Chatterjee}}, \bibinfo {author} {\bibfnamefont {A.}~\bibnamefont
  {Annamareddy}}, \bibinfo {author} {\bibfnamefont {J.}~\bibnamefont
  {Ketkaew}}, \bibinfo {author} {\bibfnamefont {J.}~\bibnamefont {Schroers}},
  \bibinfo {author} {\bibfnamefont {D.}~\bibnamefont {Morgan}},\ and\ \bibinfo
  {author} {\bibfnamefont {P.~M.}\ \bibnamefont {Voyles}},\ }\bibfield  {title}
  {\bibinfo {title} {{Fast Surface Dynamics on a Metallic Glass Nanowire}},\
  }\href {https://doi.org/10.1021/acsnano.1c00500} {\bibfield  {journal}
  {\bibinfo  {journal} {ACS Nano}\ }\textbf {\bibinfo {volume} {15}},\ \bibinfo
  {pages} {11309} (\bibinfo {year} {2021})}\BibitemShut {NoStop}%
\bibitem [{\citenamefont {Qiao}\ \emph {et~al.}(2014)\citenamefont {Qiao},
  \citenamefont {Casalini}, \citenamefont {Pelletier},\ and\ \citenamefont
  {Kato}}]{Qiao2014}%
  \BibitemOpen
  \bibfield  {author} {\bibinfo {author} {\bibfnamefont {J.}~\bibnamefont
  {Qiao}}, \bibinfo {author} {\bibfnamefont {R.}~\bibnamefont {Casalini}},
  \bibinfo {author} {\bibfnamefont {J.-M.}\ \bibnamefont {Pelletier}},\ and\
  \bibinfo {author} {\bibfnamefont {H.}~\bibnamefont {Kato}},\ }\bibfield
  {title} {\bibinfo {title} {{Characteristics of the Structural and
  Johari--Goldstein Relaxations in Pd-Based Metallic Glass-Forming Liquids}},\
  }\href {https://doi.org/10.1021/jp4121782} {\bibfield  {journal} {\bibinfo
  {journal} {J. Phys. Chem. B}\ }\textbf {\bibinfo {volume} {118}},\ \bibinfo
  {pages} {3720} (\bibinfo {year} {2014})}\BibitemShut {NoStop}%
\bibitem [{\citenamefont {Yu}\ \emph {et~al.}(2013)\citenamefont {Yu},
  \citenamefont {Wang},\ and\ \citenamefont {Samwer}}]{Yu2013}%
  \BibitemOpen
  \bibfield  {author} {\bibinfo {author} {\bibfnamefont {H.}~\bibnamefont
  {Yu}}, \bibinfo {author} {\bibfnamefont {W.~H.}\ \bibnamefont {Wang}},\ and\
  \bibinfo {author} {\bibfnamefont {K.}~\bibnamefont {Samwer}},\ }\bibfield
  {title} {\bibinfo {title} {{The {$\beta$} relaxation in metallic glasses: An
  overview}},\ }\href {https://doi.org/10.1016/j.mattod.2013.05.002} {\bibfield
   {journal} {\bibinfo  {journal} {Mater. Today}\ }\textbf {\bibinfo {volume}
  {16}},\ \bibinfo {pages} {183} (\bibinfo {year} {2013})}\BibitemShut
  {NoStop}%
\bibitem [{\citenamefont {Liu}\ \emph {et~al.}(2016)\citenamefont {Liu},
  \citenamefont {Pineda}, \citenamefont {Qiao},\ and\ \citenamefont
  {Crespo}}]{Liu2016}%
  \BibitemOpen
  \bibfield  {author} {\bibinfo {author} {\bibfnamefont {C.}~\bibnamefont
  {Liu}}, \bibinfo {author} {\bibfnamefont {E.}~\bibnamefont {Pineda}},
  \bibinfo {author} {\bibfnamefont {J.}~\bibnamefont {Qiao}},\ and\ \bibinfo
  {author} {\bibfnamefont {D.}~\bibnamefont {Crespo}},\ }\bibfield  {title}
  {\bibinfo {title} {{Modeling of the Sub-$T_g$ Relaxation Spectrum of
  Pd$_{42.5}$Ni$_{7.5}$Cu$_{30}$P$_{20}$ Metallic Glass}},\ }\href
  {https://doi.org/10.1021/acs.jpcb.5b11754} {\bibfield  {journal} {\bibinfo
  {journal} {J. Phys. Chem. B}\ }\textbf {\bibinfo {volume} {120}},\ \bibinfo
  {pages} {2838} (\bibinfo {year} {2016})}\BibitemShut {NoStop}%
\bibitem [{\citenamefont {Hirata}\ and\ \citenamefont
  {Chen}(2014)}]{Hirata2014}%
  \BibitemOpen
  \bibfield  {author} {\bibinfo {author} {\bibfnamefont {A.}~\bibnamefont
  {Hirata}}\ and\ \bibinfo {author} {\bibfnamefont {M.}~\bibnamefont {Chen}},\
  }\bibfield  {title} {\bibinfo {title} {{Angstrom-beam electron diffraction of
  amorphous materials}},\ }\href
  {https://doi.org/10.1016/j.jnoncrysol.2013.03.010} {\bibfield  {journal}
  {\bibinfo  {journal} {J. Non-Cryst. Solids}\ }\textbf {\bibinfo {volume}
  {383}},\ \bibinfo {pages} {52} (\bibinfo {year} {2014})}\BibitemShut
  {NoStop}%
\bibitem [{\citenamefont {Hirata}(2021)}]{Hirata2021}%
  \BibitemOpen
  \bibfield  {author} {\bibinfo {author} {\bibfnamefont {A.}~\bibnamefont
  {Hirata}},\ }\bibfield  {title} {\bibinfo {title} {{Local structure analysis
  of amorphous materials by angstrom-beam electron diffraction}},\ }\href
  {https://doi.org/10.1093/jmicro/dfaa075} {\bibfield  {journal} {\bibinfo
  {journal} {Microscopy}\ }\textbf {\bibinfo {volume} {70}},\ \bibinfo {pages}
  {171} (\bibinfo {year} {2021})}\BibitemShut {NoStop}%
\bibitem [{\citenamefont {Huang}\ \emph {et~al.}(2022)\citenamefont {Huang},
  \citenamefont {Francis}, \citenamefont {Ketkaew}, \citenamefont {Schroers},\
  and\ \citenamefont {Voyles}}]{Huang2022}%
  \BibitemOpen
  \bibfield  {author} {\bibinfo {author} {\bibfnamefont {S.}~\bibnamefont
  {Huang}}, \bibinfo {author} {\bibfnamefont {C.}~\bibnamefont {Francis}},
  \bibinfo {author} {\bibfnamefont {J.}~\bibnamefont {Ketkaew}}, \bibinfo
  {author} {\bibfnamefont {J.}~\bibnamefont {Schroers}},\ and\ \bibinfo
  {author} {\bibfnamefont {P.~M.}\ \bibnamefont {Voyles}},\ }\bibfield  {title}
  {\bibinfo {title} {{Correlation symmetry analysis of electron nanodiffraction
  from amorphous materials}},\ }\href
  {https://doi.org/10.1016/j.ultramic.2021.113405} {\bibfield  {journal}
  {\bibinfo  {journal} {Ultramicroscopy}\ }\textbf {\bibinfo {volume} {232}},\
  \bibinfo {pages} {113405} (\bibinfo {year} {2022})}\BibitemShut {NoStop}%
\end{thebibliography}%


\begin{thebibliography}{6}%
\makeatletter
\providecommand \@ifxundefined [1]{%
 \@ifx{#1\undefined}
}%
\providecommand \@ifnum [1]{%
 \ifnum #1\expandafter \@firstoftwo
 \else \expandafter \@secondoftwo
 \fi
}%
\providecommand \@ifx [1]{%
 \ifx #1\expandafter \@firstoftwo
 \else \expandafter \@secondoftwo
 \fi
}%
\providecommand \natexlab [1]{#1}%
\providecommand \enquote  [1]{``#1''}%
\providecommand \bibnamefont  [1]{#1}%
\providecommand \bibfnamefont [1]{#1}%
\providecommand \citenamefont [1]{#1}%
\providecommand \href@noop [0]{\@secondoftwo}%
\providecommand \href [0]{\begingroup \@sanitize@url \@href}%
\providecommand \@href[1]{\@@startlink{#1}\@@href}%
\providecommand \@@href[1]{\endgroup#1\@@endlink}%
\providecommand \@sanitize@url [0]{\catcode `\\12\catcode `\$12\catcode
  `\&12\catcode `\#12\catcode `\^12\catcode `\_12\catcode `\%12\relax}%
\providecommand \@@startlink[1]{}%
\providecommand \@@endlink[0]{}%
\providecommand \url  [0]{\begingroup\@sanitize@url \@url }%
\providecommand \@url [1]{\endgroup\@href {#1}{\urlprefix }}%
\providecommand \urlprefix  [0]{URL }%
\providecommand \Eprint [0]{\href }%
\providecommand \doibase [0]{https://doi.org/}%
\providecommand \selectlanguage [0]{\@gobble}%
\providecommand \bibinfo  [0]{\@secondoftwo}%
\providecommand \bibfield  [0]{\@secondoftwo}%
\providecommand \translation [1]{[#1]}%
\providecommand \BibitemOpen [0]{}%
\providecommand \bibitemStop [0]{}%
\providecommand \bibitemNoStop [0]{.\EOS\space}%
\providecommand \EOS [0]{\spacefactor3000\relax}%
\providecommand \BibitemShut  [1]{\csname bibitem#1\endcsname}%
\let\auto@bib@innerbib\@empty
\bibitem [{\citenamefont {Egerton}(2011)}]{Egerton2011}%
  \BibitemOpen
  \bibfield  {author} {\bibinfo {author} {\bibfnamefont {R.~F.}\ \bibnamefont
  {Egerton}},\ }\href {https://doi.org/10.1007/978-1-4419-9583-4} {\emph
  {\bibinfo {title} {{Electron Energy-Loss Spectroscopy in the Electron
  Microscope}}}},\ \bibinfo {edition} {3rd}\ ed.\ (\bibinfo  {publisher}
  {Springer},\ \bibinfo {address} {New York},\ \bibinfo {year}
  {2011})\BibitemShut {NoStop}%
\bibitem [{\citenamefont {Chatterjee}\ \emph {et~al.}(2021)\citenamefont
  {Chatterjee}, \citenamefont {Annamareddy}, \citenamefont {Ketkaew},
  \citenamefont {Schroers}, \citenamefont {Morgan},\ and\ \citenamefont
  {Voyles}}]{Chatterjee2021}%
  \BibitemOpen
  \bibfield  {author} {\bibinfo {author} {\bibfnamefont {D.}~\bibnamefont
  {Chatterjee}}, \bibinfo {author} {\bibfnamefont {A.}~\bibnamefont
  {Annamareddy}}, \bibinfo {author} {\bibfnamefont {J.}~\bibnamefont
  {Ketkaew}}, \bibinfo {author} {\bibfnamefont {J.}~\bibnamefont {Schroers}},
  \bibinfo {author} {\bibfnamefont {D.}~\bibnamefont {Morgan}},\ and\ \bibinfo
  {author} {\bibfnamefont {P.~M.}\ \bibnamefont {Voyles}},\ }\bibfield  {title}
  {\bibinfo {title} {{Fast Surface Dynamics on a Metallic Glass Nanowire}},\
  }\href {https://doi.org/10.1021/acsnano.1c00500} {\bibfield  {journal}
  {\bibinfo  {journal} {ACS Nano}\ }\textbf {\bibinfo {volume} {15}},\ \bibinfo
  {pages} {11309} (\bibinfo {year} {2021})}\BibitemShut {NoStop}%
\bibitem [{\citenamefont {Spangenberg}\ \emph {et~al.}(2021)\citenamefont
  {Spangenberg}, \citenamefont {Hilke}, \citenamefont {Wilde},\ and\
  \citenamefont {Peterlechner}}]{Spangenberg2021}%
  \BibitemOpen
  \bibfield  {author} {\bibinfo {author} {\bibfnamefont {K.}~\bibnamefont
  {Spangenberg}}, \bibinfo {author} {\bibfnamefont {S.}~\bibnamefont {Hilke}},
  \bibinfo {author} {\bibfnamefont {G.}~\bibnamefont {Wilde}},\ and\ \bibinfo
  {author} {\bibfnamefont {M.}~\bibnamefont {Peterlechner}},\ }\bibfield
  {title} {\bibinfo {title} {{Direct View on Non-Equilibrium Heterogeneous
  Dynamics in Glassy Nanorods}},\ }\href
  {https://doi.org/10.1002/adfm.202103742} {\bibfield  {journal} {\bibinfo
  {journal} {Adv. Funct. Mater.}\ }\textbf {\bibinfo {volume} {31}},\ \bibinfo
  {pages} {2103742} (\bibinfo {year} {2021})}\BibitemShut {NoStop}%
\bibitem [{\citenamefont {Zhang}\ \emph {et~al.}(2018)\citenamefont {Zhang},
  \citenamefont {Maldonis}, \citenamefont {Liu}, \citenamefont {Schroers},\
  and\ \citenamefont {Voyles}}]{Zhang2018}%
  \BibitemOpen
  \bibfield  {author} {\bibinfo {author} {\bibfnamefont {P.}~\bibnamefont
  {Zhang}}, \bibinfo {author} {\bibfnamefont {J.~J.}\ \bibnamefont {Maldonis}},
  \bibinfo {author} {\bibfnamefont {Z.}~\bibnamefont {Liu}}, \bibinfo {author}
  {\bibfnamefont {J.}~\bibnamefont {Schroers}},\ and\ \bibinfo {author}
  {\bibfnamefont {P.~M.}\ \bibnamefont {Voyles}},\ }\bibfield  {title}
  {\bibinfo {title} {{Spatially heterogeneous dynamics in a metallic glass
  forming liquid imaged by electron correlation microscopy}},\ }\href
  {https://doi.org/10.1038/s41467-018-03604-2} {\bibfield  {journal} {\bibinfo
  {journal} {Nat. Commun.}\ }\textbf {\bibinfo {volume} {9}},\ \bibinfo {pages}
  {1129} (\bibinfo {year} {2018})}\BibitemShut {NoStop}%
\bibitem [{\citenamefont {Nakazawa}\ \emph {et~al.}(2024)\citenamefont
  {Nakazawa}, \citenamefont {Mitsuishi}, \citenamefont {Iakoubovskii},
  \citenamefont {Kohara},\ and\ \citenamefont {Tsuchiya}}]{Nakazawa2024}%
  \BibitemOpen
  \bibfield  {author} {\bibinfo {author} {\bibfnamefont {K.}~\bibnamefont
  {Nakazawa}}, \bibinfo {author} {\bibfnamefont {K.}~\bibnamefont {Mitsuishi}},
  \bibinfo {author} {\bibfnamefont {K.}~\bibnamefont {Iakoubovskii}}, \bibinfo
  {author} {\bibfnamefont {S.}~\bibnamefont {Kohara}},\ and\ \bibinfo {author}
  {\bibfnamefont {K.}~\bibnamefont {Tsuchiya}},\ }\bibfield  {title} {\bibinfo
  {title} {{Structure-dynamics relation in metallic glass revealed by
  5-dimensional scanning transmission electron microscopy}},\ }\href
  {https://doi.org/10.1038/s41427-024-00577-1} {\bibfield  {journal} {\bibinfo
  {journal} {NPG Asia Mater.}\ }\textbf {\bibinfo {volume} {16}},\ \bibinfo
  {pages} {57} (\bibinfo {year} {2024})}\BibitemShut {NoStop}%
\bibitem [{\citenamefont {Nakazawa}\ and\ \citenamefont
  {Mitsuishi}(2025)}]{Nakazawa2025}%
  \BibitemOpen
  \bibfield  {author} {\bibinfo {author} {\bibfnamefont {K.}~\bibnamefont
  {Nakazawa}}\ and\ \bibinfo {author} {\bibfnamefont {K.}~\bibnamefont
  {Mitsuishi}},\ }\bibfield  {title} {\bibinfo {title} {{Relaxation time
  measurement: correlating diffraction patterns}},\ }\href
  {https://doi.org/10.1093/jmicro/dfaf048} {\bibfield  {journal} {\bibinfo
  {journal} {Microscopy}\ }\textbf {\bibinfo {volume} {75}},\ \bibinfo {pages}
  {285} (\bibinfo {year} {2025})}\BibitemShut {NoStop}%
\end{thebibliography}%

\end{document}


\title{Supplemental Material for\\
Nanoscale two-component structural relaxation in the fragile metallic glass \PNCP\ revealed by
five-dimensional scanning transmission electron microscopy}

\author{Katsuaki Nakazawa}
\affiliation{Center for Basic Research on Materials, National Institute for Materials Science,
1-2-1 Sengen, Tsukuba, Ibaraki 305-0047, Japan}
\author{Hidemi Kato}
\author{Tetsu Ichitsubo}
\affiliation{Institute for Materials Research, Tohoku University, 2-1-1 Aoba-ku, Sendai, Miyagi 980-8577, Japan}
\author{Shinji Kohara}
\author{Kazutaka Mitsuishi}
\affiliation{Center for Basic Research on Materials, National Institute for Materials Science,
1-2-1 Sengen, Tsukuba, Ibaraki 305-0047, Japan}
\author{Koichi Tsuchiya}
\affiliation{Research Center for Structural Materials, National Institute for Materials Science,
Tsukuba, Ibaraki 305-0047, Japan}

\date{\today}
\maketitle

\section{Sample thickness}
\label{sec:thick}

\begin{figure}[tbp]
\centering
\includegraphics[width=\linewidth]{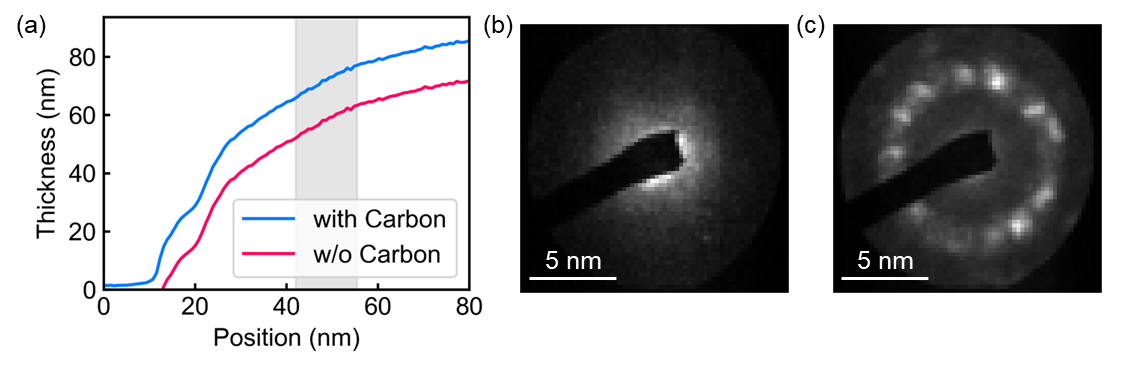}
\caption{(a) Thickness profile measured by electron energy loss spectroscopy. The blue line is the
thickness calculated without the carbon layer and the red line with the carbon layer taken into account;
the gray region marks the area used for the analysis. (b) Temporal average of the diffraction pattern
from the carbon layer alone and (c) from the analyzed area, both at 533~K.}
\label{fig:thick}
\end{figure}

The thickness was measured by electron energy loss spectroscopy~\cite{Egerton2011}. The inelastic mean
free paths of carbon and of PNCP are 160~nm~\cite{Egerton2011} and 110~nm, respectively. Each surface of
the sample was covered with 10~nm of amorphous carbon to suppress surface diffusion~\cite{Chatterjee2021},
which accelerates the dynamics in thin regions and shortens the apparent relaxation time. The thickness
profile is shown in Fig.~\ref{fig:thick}(a); the thickness of PNCP in the analyzed area was 50--60~nm,
a value comparable to those used in previous work~\cite{Spangenberg2021,Zhang2018,Nakazawa2024,Nakazawa2025}.
Figures~\ref{fig:thick}(b) and \ref{fig:thick}(c) show the temporal averages of the diffraction patterns
obtained from the carbon layer alone and from the analyzed area. Carbon scatters few electrons in the
range of spatial frequency where PNCP forms the halo and the speckles (4.4--4.6~nm$^{-1}$), so the carbon
layer does not affect the present analysis.

\begin{figure}[tbp]
\centering
\includegraphics[width=\linewidth]{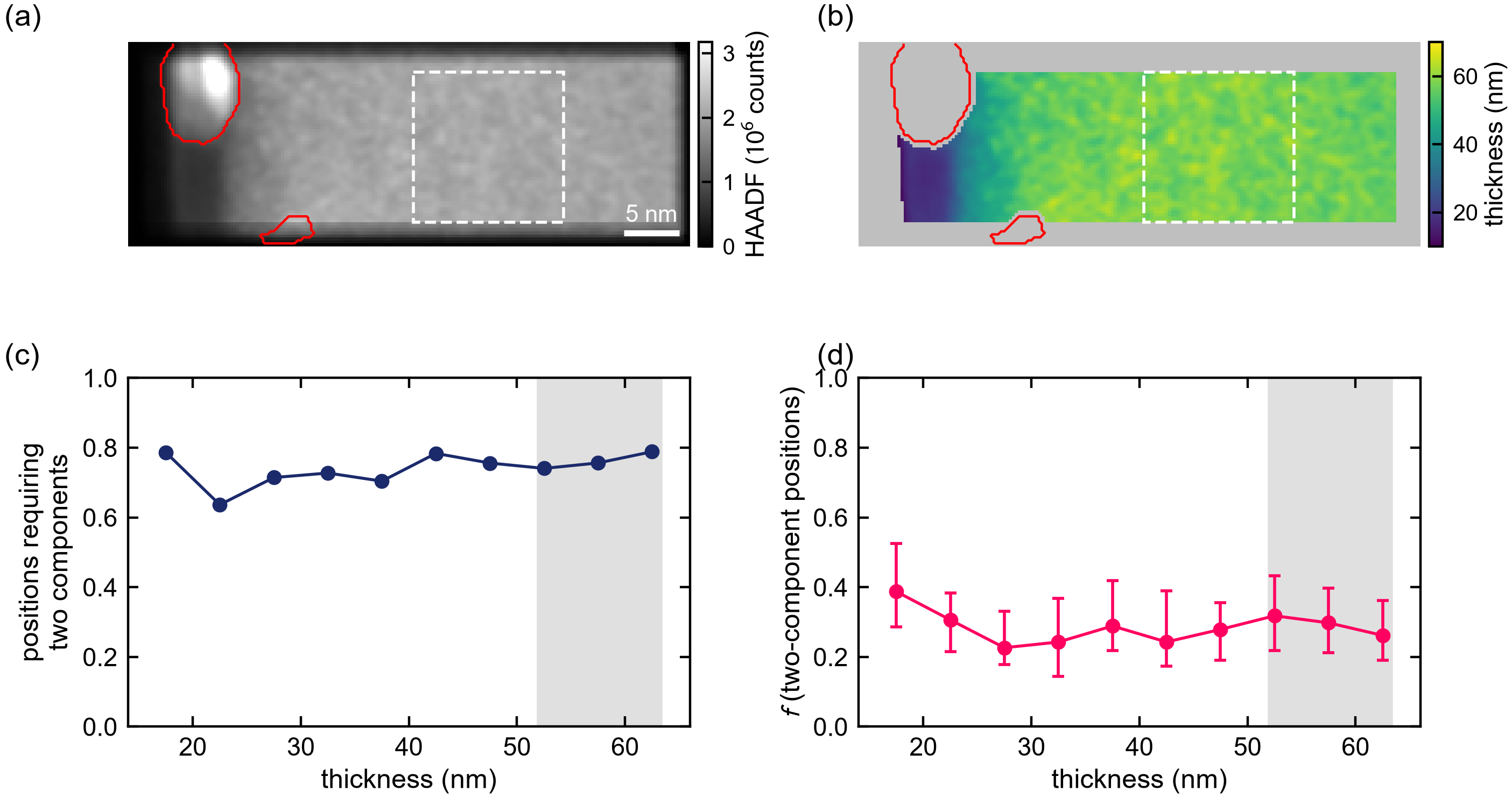}
\caption{Thickness dependence at 533~K (reference). (a) HAADF image averaged over the measurement and (b)
thickness estimated from it; positions excluded from the analysis are shown in gray in (b). Red lines
enclose the crystallized regions identified from the fluctuation, and the dashed rectangle marks the
analyzed area. The scale bar is 5~nm. (c) Fraction of positions requiring two components and (d) median
$f$ at these positions as a function of the thickness, in bins of 5~nm; bars in (d) give the interquartile
range. The gray band marks the thickness range of the analyzed area according to Fig.~\ref{fig:thick}(a).}
\label{fig:thickdep}
\end{figure}

For reference, we examined whether the fraction of positions requiring two components and $f$ depend on
the thickness, using the data at 533~K (PNCP\_003), which are not used in the main text because the slow
component is not resolved within the observation time. At this temperature the field of view contains a
thin amorphous region next to the edge of the sample [Fig.~\ref{fig:thickdep}(a)]; at higher
temperatures this thin region has crystallized, so the thickness dependence can be examined only at
533~K. The local thickness was
estimated from the high-angle annular dark-field (HAADF) intensity averaged over the measurement,
assuming that the intensity is proportional to the thickness (the intensity in vacuum is 0.4\% of that in
the analyzed area); the proportionality constant was fixed so that the median intensity in the analyzed
area corresponds to the mean thickness of the gray region in Fig.~\ref{fig:thick}(a), 58~nm
[Fig.~\ref{fig:thickdep}(b)]. Crystallized regions were identified from the time series of the
fluctuation, as positions where $V$ exceeds the median of the analyzed area by more than six times the
median absolute deviation in more than 5\% of the frames, and were excluded together with a margin of two
positions. The same model selection as in the main text was applied to all positions.

Between 15 and 65~nm, the fraction of positions requiring two components stays at 0.64--0.79 and the
median $f$ at these positions at 0.23--0.39, without a systematic trend [Figs.~\ref{fig:thickdep}(c) and
(d)]. If the fast component arose from a surface layer of fixed thickness, $f$ would scale inversely with
the thickness and would increase from 0.3 at 58~nm to about 0.6 at 30~nm; this is not observed. If the two
components arose from the stacking of regions with different dynamics along the beam direction, the
fraction of two-component positions and the spread of $f$ would change with the number of stacked regions,
that is, with the thickness; neither changes appreciably over a threefold range of thickness. These data
therefore favor a two-component relaxation that occurs locally within the probed volume. We regard this
result as supporting evidence only: at 533~K the slow component (median $\tauslow \approx 1.5$--$2.5
\times 10^{3}$~s) is longer than the observation time (about 550~s), the false-positive rate of the model
selection was not evaluated at this temperature, the thickness relies on the proportionality between the
HAADF intensity and the thickness, and the spatial autocorrelation was not taken into account.

\FloatBarrier
\section{Model selection and the reliability of the fitted parameters}
\label{sec:model}

At every probe position the normalized correlation function was fitted with a one-component KWW function
(free parameters $\tau$, $\gamma$, the amplitude $A$ and a baseline $B$ restricted to $|B| \le 0.1$) and
with the sum of two KWW functions (free parameters $\taufast$, $\gammafast$, $\tauslow$, $\gammaslow$,
$A$ and $f$, with $B = 0$). The baseline of the one-component KWW function absorbs the noise floor that remains
after the relaxation is complete; no baseline is used in the two-component function because a slow
component with $\tauslow$ much longer than the observation time is indistinguishable from a constant.
The two models are therefore not nested, and the Bayesian information criterion (BIC) is used as the
primary criterion; the nested $F$-test is quoted for reference only.

The fit residuals are serially correlated (lag-1 autocorrelation $\rho_1 \approx 0.24$--$0.67$), so
treating the $n \approx 288$ (558 and 563~K) or 144 (568--578~K) data points as independent would
overestimate the evidence for the second component. The BIC was therefore evaluated with the effective
number of data points $n_{\mathrm{eff}} = n(1-\rho_1)/(1+\rho_1) \approx 48$--$89$, where $\rho_1$ is the
lag-1 autocorrelation of the residuals of the two-component fit, evaluated at each position. Without this
correction the fraction of two-component positions at 558~K would be 0.81 instead of 0.61. Table~\ref{tab:sel} lists, for each temperature, the fraction of
positions at which two components are supported, together with the same fraction obtained with the
nested $F$-test and without the autocorrelation correction, and the false-positive rates obtained
from synthetic one-component data with the measured noise.

\begin{table}[tbp]
\caption{Fraction of positions at which two components are required, and the corresponding
false-positive rates obtained from synthetic one-component data (600 realizations per temperature).
$N = 2500$ positions at each temperature.}
\label{tab:sel}
\begin{ruledtabular}
\begin{tabular}{ccccccc}
 & \multicolumn{3}{c}{measured fraction} & \multicolumn{2}{c}{residuals}
 & false positives\\
\cline{2-4}\cline{5-6}
$T$ (K) & BIC & $F$-test & BIC (uncorr.) & $n_{\mathrm{eff}}$ & $\rho_1$
 & (BIC, \%) \\
\hline
558 & 0.61 & 0.65 & 0.81 & 63 & 0.64 & 0.0\\
563 & 0.27 & 0.31 & 0.47 & 56 & 0.67 & 0.2\\
568 & 0.11 & 0.14 & 0.26 & 48 & 0.50 & 0.5\\
573 & 0.05 & 0.08 & 0.14 & 59 & 0.41 & 0.0\\
578 & 0.02 & 0.04 & 0.04 & 89 & 0.24 & 0.0\\
\end{tabular}
\end{ruledtabular}
\end{table}

In the test with 300 frames described in the main text (Sec.~III~C), the fraction of two-component
positions is 0.55 for the first and 0.56 for the last 300 frames at 558~K, and 0.39 and 0.38 at 563~K.

At the two shortest delay times the two-component fit also leaves a positive mean residual ($+0.04$ and
$+0.02$ at 558~K), because the correlation decays more slowly there than either model predicts. To test
whether these points affect the analysis, the 558~K data were re-fitted without the first data points.
The fraction of two-component positions is 0.62 with all points, 0.62 without the first two, and 0.61 without the first
four, compared with 0.61 in the main analysis, so the classification is unchanged. The median $\taufast$
at the two-component positions increases from 100 to 126~s when the first two points are excluded; we
therefore regard $\taufast$ as accurate to about 30\%.

\begin{figure}[tbp]
\centering
\includegraphics[width=\linewidth]{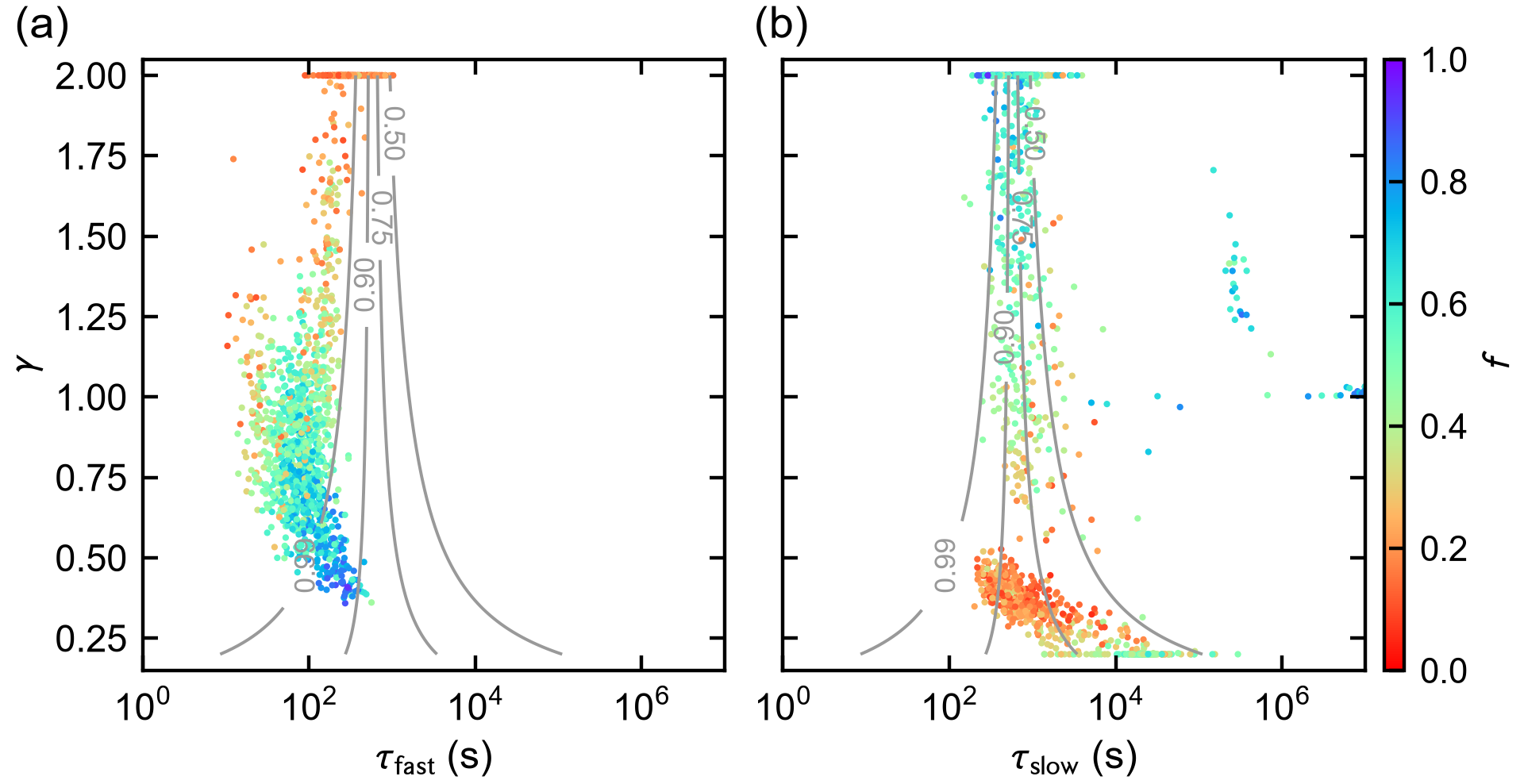}
\caption{Fitted (a) $(\taufast, \gammafast)$ and (b) $(\tauslow, \gammaslow)$ at the two-component
positions at 558~K, colored by $f$. Gray contours show the fraction of the corresponding component that
has decayed within the longest delay time used in the fit (554~s).}
\label{fig:reliab}
\end{figure}

The reliability of the fitted relaxation times is limited by the observation time $t_{\mathrm{obs}}$,
the longest delay time used in the fit, which is half the duration of the measurement (554~s at 558 and
563~K and 277~s at 568--578~K). The fraction of the relaxation that is completed within $t_{\mathrm{obs}}$ is $1-\exp[-2(t_{\mathrm{obs}}/\tau)^{\gamma}]$, and is drawn as
contours in Fig.~\ref{fig:reliab}. Most of the $(\taufast, \gammafast)$ values lie where more than 90\%
of the fast component has decayed within $t_{\mathrm{obs}}$. In contrast, a separate group of positions has
$\tauslow > 10^{5}$~s; there the slow component barely
changes during the measurement and acts as a constant background. These positions (4.0\% of the
two-component positions at 558~K) were excluded from the averages of $\tauslow$ in the main text.

\FloatBarrier
\section{Why the rank correlations between structure and dynamics are small}
\label{sec:corr}

The rank correlations between the relaxation and the structural parameters are significant but small
($|\rho| \le 0.21$). Two effects reduce them: the noise in the position-wise $f$ (Sec.~\ref{sec:corrA}) and the change of
the structure during the measurement (Secs.~\ref{sec:corrB} and \ref{sec:corrC}). Both effects weaken
the observed correlations, so that the underlying correlation is not directly given by the observed values.

\subsection{Noise in $f$}
\label{sec:corrA}

The position-wise $f$ is noisier than the structural parameters. Among the two-component positions, the
nearest-neighbor spatial correlation of $f$ is 0.40 and 0.24 at 558 and 563~K, whereas it is 0.63 and 0.80
for $\log \tauslow$ and 0.83 and 0.73 for the fluctuation; the structural parameters are in addition averaged over the whole measurement. Measurement
noise in one of the two variables attenuates a rank correlation, so the observed values are lower bounds
on the underlying correlation. Part of the low spatial correlation of $f$ may, however, reflect the
change of the $f$ map during the measurement rather than noise (Sec.~\ref{sec:corrC}).

\subsection{Temporal change of the structure}
\label{sec:corrB}

\begin{figure}[tbp]
\centering
\includegraphics[width=\linewidth]{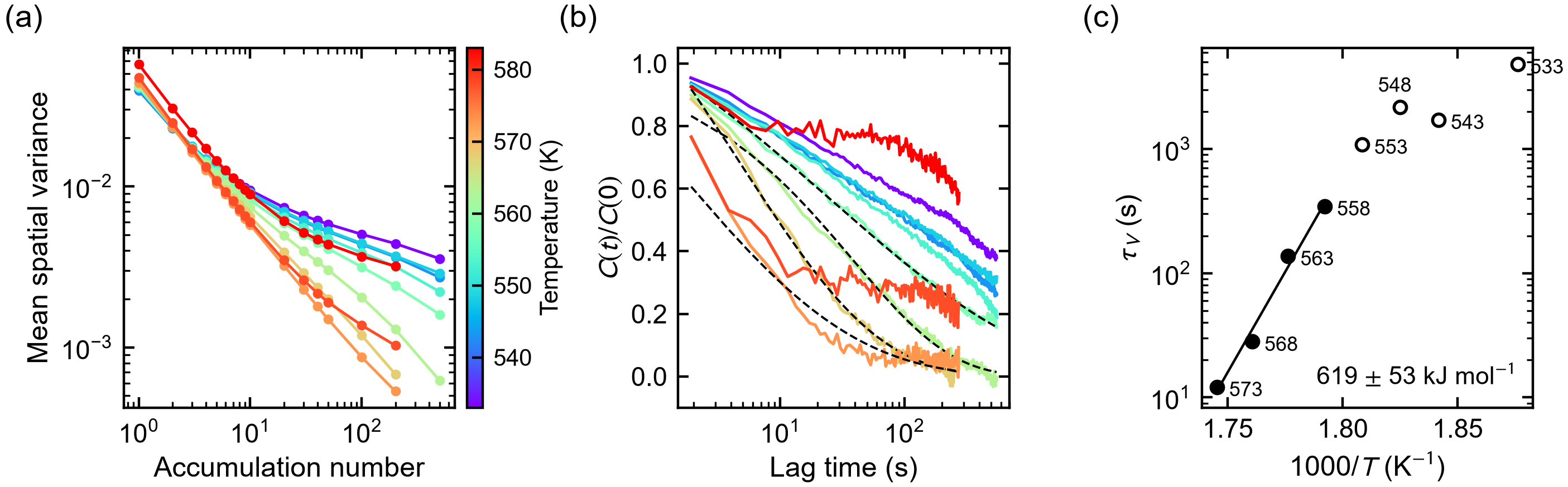}
\caption{(a) Mean spatial variance of the time-averaged fluctuation as a function of the averaging window
(accumulation number, in frames) at each temperature (color). The fluctuation $V$ of each frame was
averaged over the window at each position, the variance of the resulting map over the positions in the
analyzed region was calculated, and the variances were averaged over the positions of the window shifted
along the time series. The longest window is 500 frames for the measurements with 600 frames and 200 frames for
those with 300 frames. (b) Correlation $C(t)$ between the fluctuation maps of two frames as a function of the lag time,
normalized by its value extrapolated to $t = 0$, at 533--583~K (colors as in (a)). Dashed lines are
one-component KWW fits without a baseline at 558--573~K. (c) Structural correlation time $\tau_V$ from
these fits at 533--573~K. The line is an Arrhenius fit to 558--573~K (filled symbols), the temperature
range of the relaxation analysis.}
\label{fig:naccum}
\end{figure}

The second effect is that the structure at each position changes during the measurement. This is seen directly in the
dependence of the spatial variance of the time-averaged fluctuation on the averaging window
(Fig.~\ref{fig:naccum}). For short windows the variance decreases steeply and is nearly independent of
temperature, as expected from the noise in the fluctuation of single frames. If the structure were static,
the variance would level off at a value set by the spatial heterogeneity of the structure; if the
structure rearranges within the window, the heterogeneity is averaged out and the variance keeps
decreasing. The variance for long windows decreases with increasing temperature, which shows that the
fluctuation changes faster in time at higher temperatures [Fig.~\ref{fig:naccum}(a)]. The curves at 578
and 583~K, which lie above that at 573~K for long windows, are affected by the partial crystallization of
the sample at these temperatures: the crystallized regions add a static contribution to the variance
(Sec.~\ref{sec:temp}).

The time scale of this change was obtained from the correlation between the fluctuation maps of two
frames. Let $C(t)$ be the spatial covariance, over the analyzed region, between the fluctuation maps of
two frames separated by the lag time $t$, after subtraction of the spatial mean of each frame. The
variance for a window of $N$ frames is the average of $C$ over all pairs of frames in the window,
$N^{-2}\sum_{i,j} C(|i-j|\,\Delta t)$, where $\Delta t$ is the frame interval. The noise of single frames,
which is independent between frames, contributes only to $C(0)$, of which it makes up most (83--95\%);
$C(t)$ for $t \ge \Delta t$ therefore reflects the change of the structure alone. As for $g(\bm{r},t)$,
$C(t)$ was normalized by its value extrapolated linearly to $t = 0$ from the first two lag times, and the
normalized correlation up to the observation time was fitted with the one-component KWW function
$A\exp[-2(t/\tau_V)^{\gamma_V}]$ without a baseline [Fig.~\ref{fig:naccum}(b)]. The structural
correlation time $\tau_V$ decreases from 346~s at 558~K to 12~s at 573~K, with $\gamma_V = 0.27$--0.55,
and its activation energy, $619 \pm 53$~kJ\,mol$^{-1}$ [Fig.~\ref{fig:naccum}(c)], agrees with that of
$\taus$ ($604 \pm 23$~kJ\,mol$^{-1}$) and with $E_{\alpha} \approx 640$~kJ\,mol$^{-1}$. At 558~K,
$\tau_V$ is close to the median $\taus$ (335~s) and much shorter than the duration of the measurement (about
1100~s). The structure at each position thus changes during the measurement on the time scale of the
structural relaxation. A two-component KWW function improves the fit only slightly at 558--568~K. At
573~K the one-component function does not reproduce the steep initial decay and the small constant part
at long lag times, so $\tau_V$ there is less certain. At 533--553~K, $\tau_V$ (1100--4800~s) exceeds the
observation time. At 578 and 583~K, the normalized correlation levels off at about 0.3 and 0.8,
indicating a static contribution from crystallized regions. At these temperatures the sample has
partially crystallized, and the crystallized region on the left-hand side of the field of view extends
into the lower-left part of the analyzed region, where the time-averaged fluctuation is low (Sec.~\ref{sec:temp});
it covers about 7\% of the analyzed region at 578~K and 13\% at 583~K, whereas the analyzed region is
free of it up to 573~K. These two temperatures, and 588~K, where the fluctuation in about half of the analyzed region is
dominated by crystallization, are not used for $\tau_V$.

\subsection{Dependence of the correlation on the time segment}
\label{sec:corrC}

\begin{figure}[tbp]
\centering
\includegraphics[width=\linewidth]{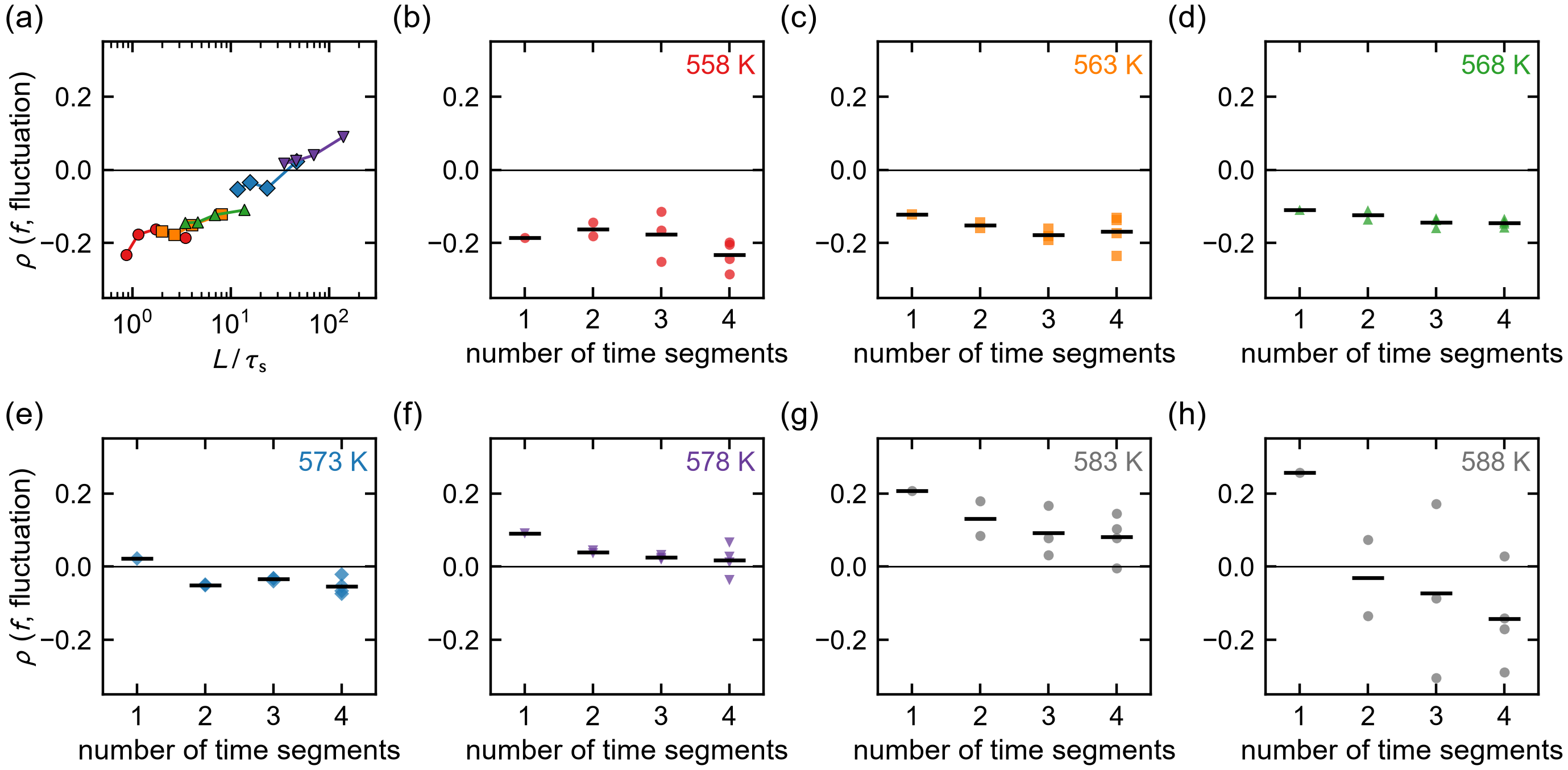}
\caption{(a) Mean Spearman correlation between $f$ and the fluctuation as a function of the segment
length $L$ divided by the median $\taus$ at each temperature, for one to four segments at 558--578~K
(colors and symbols as in (b)--(f)). (b)--(h) Dependence of the correlation on the number of time
segments at (b) 558, (c) 563, (d) 568, (e) 573, (f) 578, (g) 583, and (h) 588~K. With $n$ segments there
are $n$ values (points); the bars mark their mean. One segment corresponds to the full measurement.}
\label{fig:timesplit}
\end{figure}

Because the structure changes during the measurement (Sec.~\ref{sec:corrB}), the relaxation measured
over the full measurement reflects an average over several structural states, and over more of them at
higher temperatures. Consistent with this, the correlation
between $f$ and the fluctuation over the full measurement weakens with increasing temperature, from
$-0.19$ at 558~K to $-0.11$ at 568~K; it is not significant at 573 and 578~K, and the sign reverses above
$T_{\mathrm{g}}$ [one segment in Figs.~\ref{fig:timesplit}(b)--(h)]. To test whether this decrease
reflects the averaging rather than a weaker coupling between the structure and the relaxation, the data
were divided into two to four time segments, and the correlation was evaluated in each segment
[Figs.~\ref{fig:timesplit}(b)--(h)]. Up to 578~K the correlation becomes more negative as the segments
become shorter (for four segments, $-0.23$, $-0.17$ and $-0.15$ on average at 558, 563 and 568~K, where
it is significant in every segment, and $-0.05$ at 573~K), although shorter segments contain fewer frames
and are therefore noisier. Within one temperature the averaging cannot be removed, because even the
shortest segments are much longer than $\tau_V$ at the higher temperatures (about 140~s compared with
$\tau_V = 12$~s at 573~K). Different temperatures can, however, be compared at the same segment length
$L$ relative to the relaxation time. Plotted against $L/\taus$, with the median $\taus$ at each
temperature, the mean correlations at 558--578~K fall on a single curve [Fig.~\ref{fig:timesplit}(a)];
for example, the correlation is $-0.16$ at 558~K and $-0.17$ at 563~K for $L/\taus \approx 2$, and
$-0.12$ at both 563 and 568~K for $L/\taus \approx 7$--8. The correlation is thus set mainly by the
number of relaxation times contained in the segment, and at 558--568~K its decrease with temperature
mainly reflects the faster structural change rather than a weaker coupling. At 573~K the correlation is
somewhat weaker than this curve, and for $L/\taus \gtrsim 30$ (578~K) it becomes positive, which the
averaging, driving the correlation toward zero, does not explain; at these temperatures fewer than 5\% of
the positions require two components. At 583 and 588~K the correlation is affected by the partial
crystallization of the sample, and at 588~K the values of the individual segments scatter widely.

The same time-interval dependence, together with the reliability of the $f$ map, was examined by fitting
the two halves of the measurement separately (300 frames each at 558 and 563~K and 150 frames at 568~K). The $f$ maps of the two halves are
uncorrelated (Spearman $\rho = 0.03$, 0.02 and 0.01 at 558, 563 and 568~K, not significant in the
surrogate test). The fluctuation maps of the two halves are correlated, more strongly at 558~K
($\rho = 0.44$) than at 563 and 568~K (0.14), consistent with the shorter $\tau_V$ at the higher
temperatures (Sec.~\ref{sec:corrB}). Within each half, $f$ is correlated with the fluctuation of the same
half ($\rho = -0.11$ to $-0.18$; these are the values of Fig.~\ref{fig:timesplit} for two segments),
whereas the correlation vanishes when $f$ of one half is compared with the fluctuation of the other
($|\rho| \le 0.07$). The within-half correlations exceed the upper bound $\sqrt{r_{ff} r_{VV}}$ expected
if the two halves differed only by noise, so the disagreement between the halves reflects a real change
of the $f$ map during the measurement rather than noise alone. A classical correction for attenuation is
therefore not applicable; instead, this result shows directly that $f$ follows the structure present
during the same time interval, consistent with the temporal change of the structure (Sec.~\ref{sec:corrB}).

\subsection{Summary}

The rank correlations between $f$ and the fluctuation are reduced by the noise in $f$ and by the change of
the structure during the measurement, which occurs on the time scale of the structural relaxation and is
faster at higher temperatures. When this averaging is taken into account, $f$ follows the structure
present during the same time interval, and at 558--568~K the correlation is set mainly by the segment
length relative to the relaxation time rather than by the temperature. The underlying correlation between the
local structure and the relaxation is therefore likely to be stronger than that obtained in the present
measurement.

\FloatBarrier
\section{Temperature dependence of the relaxation and structural parameters}
\label{sec:temp}

\subsection{Relaxation parameters}

\begin{figure}[tbp]
\centering
\includegraphics[width=0.8\linewidth]{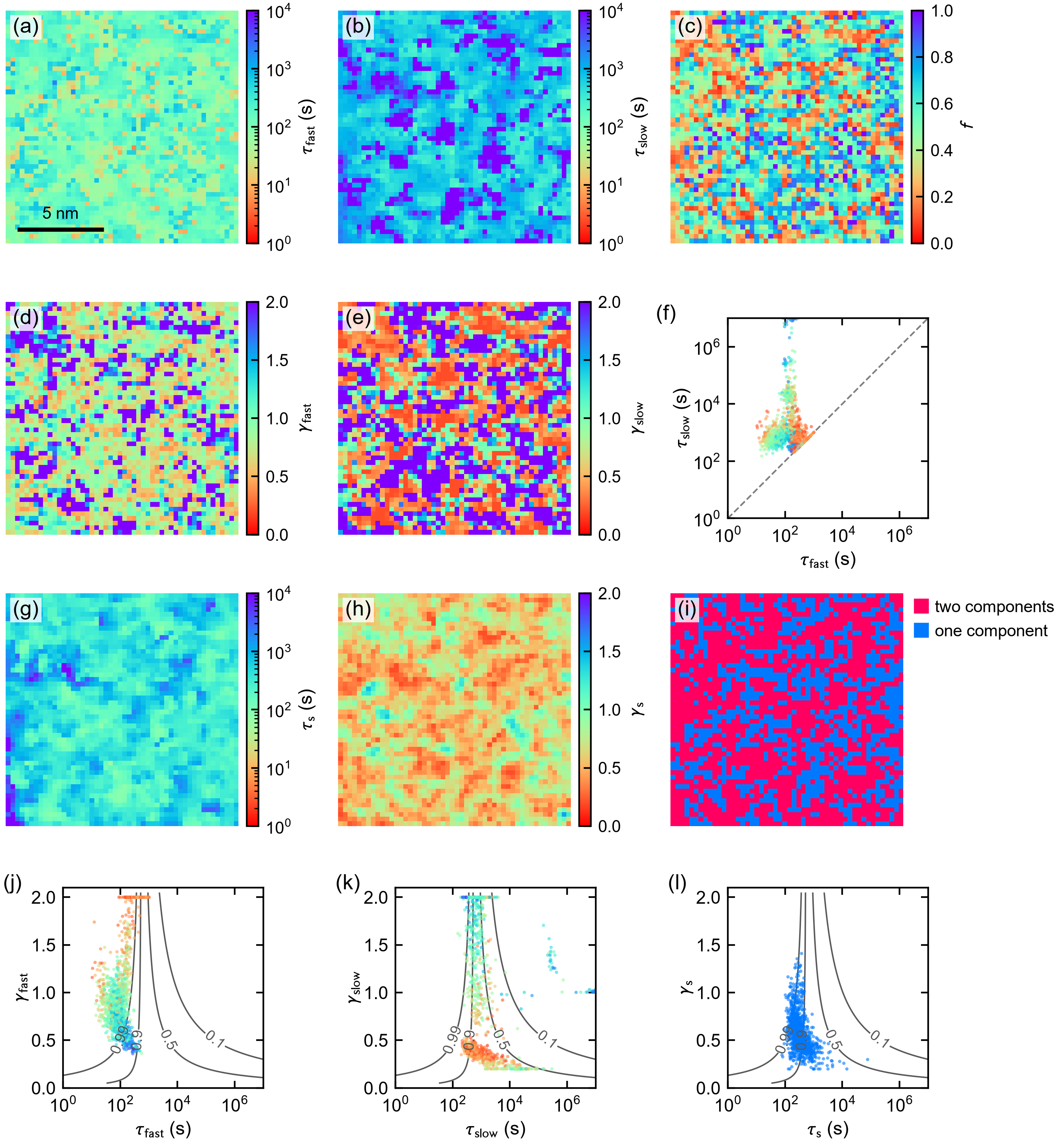}
\caption{Relaxation parameters at 558~K. The maps (a)--(e), (g) and (h) show the parameters of each fit
at every position in the field of view, regardless of the classification in (i); a parameter of a model
not selected at a position is shown for completeness only. (a)--(e) Maps of the two-component KWW fit: (a) $\tau_{\mathrm{fast}}$, (b) $\tau_{\mathrm{slow}}$, (c) $f$,
(d) $\gamma_{\mathrm{fast}}$, and (e) $\gamma_{\mathrm{slow}}$. (f) $\tau_{\mathrm{slow}}$ against
$\tau_{\mathrm{fast}}$ at the positions requiring two components; the dashed line marks $\tau_{\mathrm{fast}} = \tau_{\mathrm{slow}}$.
(g), (h) Maps of the one-component KWW fit: (g) $\tau_{\mathrm{s}}$ and (h) $\gamma_{\mathrm{s}}$.
(i) Positions requiring two components (red) and one-component positions (blue), classified by the
BIC corrected for the effective sample size. (j)--(l) Stretching exponent against relaxation time for
(j) the fast component and (k) the slow component at the positions requiring two components, colored by
$f$ with the color scale of (c), and (l) the one-component KWW fit at the one-component positions. The contours give the fraction $1-\exp[-2(T/\tau)^{\gamma}]$ that relaxes
within the measurement window $T = n\,\Delta t$ ($n \approx 288$ at 558 and 563~K and $n \approx 144$ from 568 to 583~K, and $n = 72$ at 588~K; $\Delta t = 1.92$~s).
The color scales are common to all temperatures (Figs.~\ref{fig:maps558}--\ref{fig:maps588}).}
\label{fig:maps558}
\end{figure}

\begin{figure}[tbp]
\centering
\includegraphics[width=0.82\linewidth]{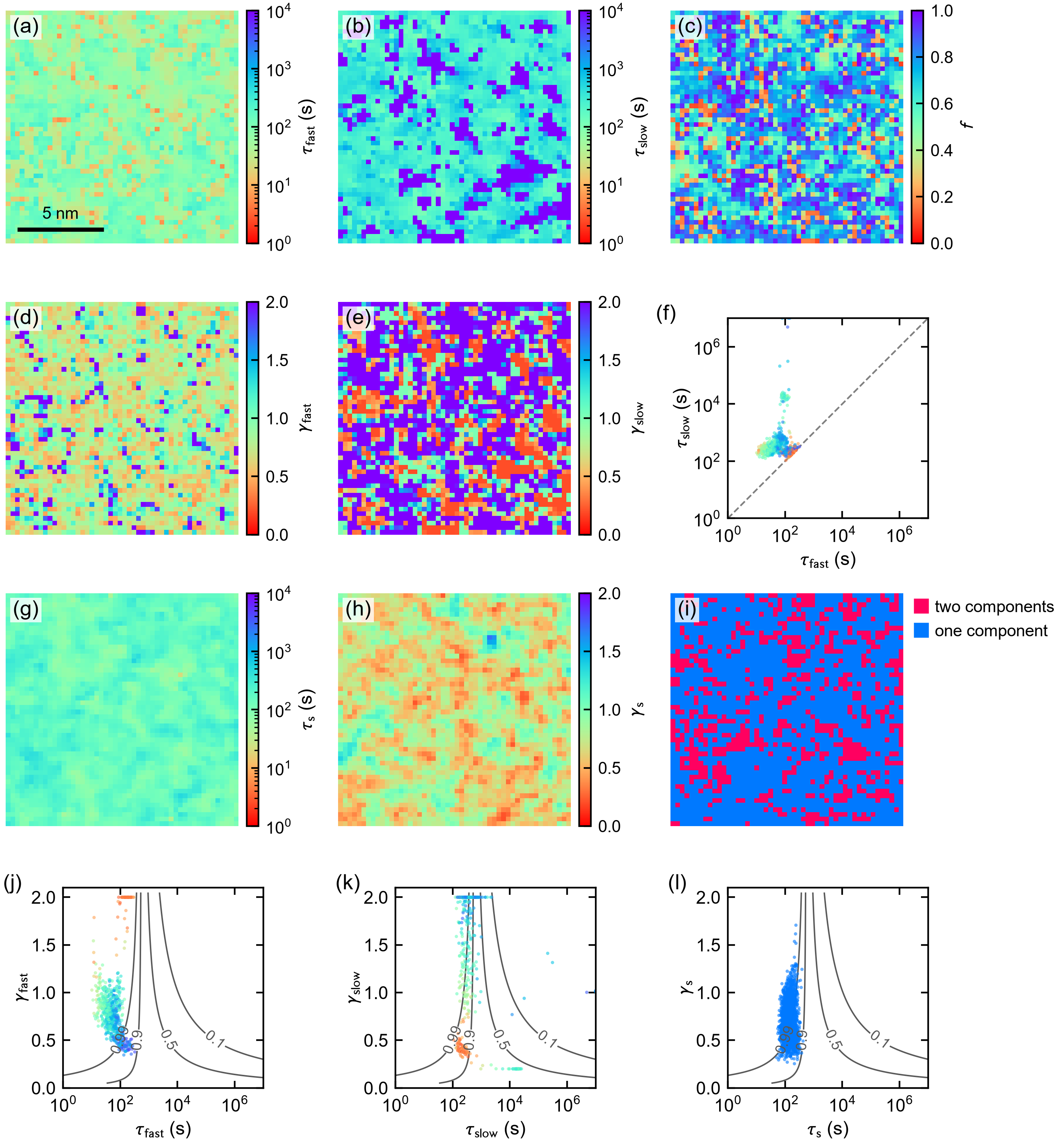}
\caption{Same as Fig.~\ref{fig:maps558} at 563~K.}
\label{fig:maps563}
\end{figure}

\begin{figure}[tbp]
\centering
\includegraphics[width=0.82\linewidth]{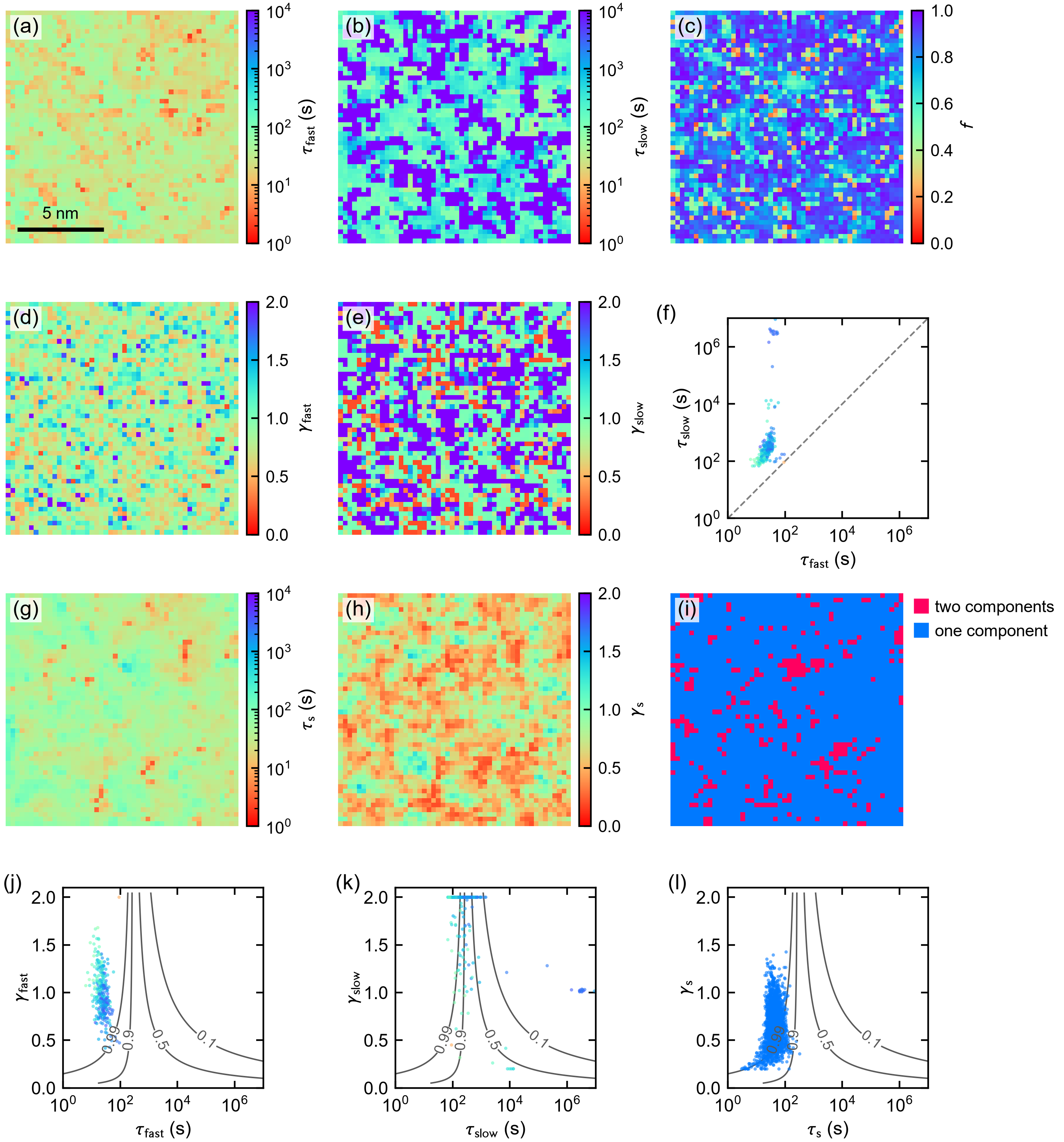}
\caption{Same as Fig.~\ref{fig:maps558} at 568~K.}
\label{fig:maps568}
\end{figure}

\begin{figure}[tbp]
\centering
\includegraphics[width=0.82\linewidth]{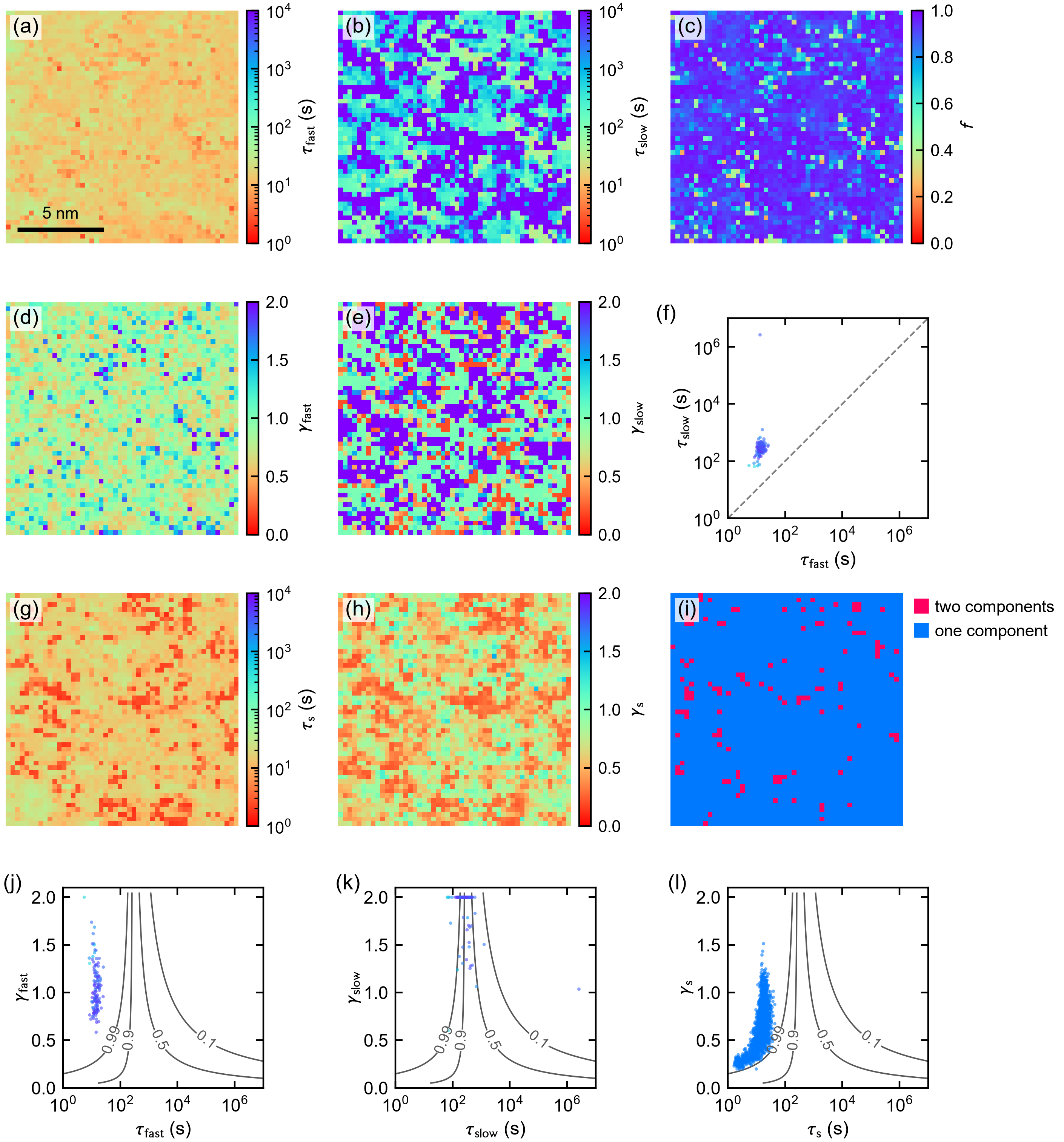}
\caption{Same as Fig.~\ref{fig:maps558} at 573~K.}
\label{fig:maps573}
\end{figure}

\begin{figure}[tbp]
\centering
\includegraphics[width=0.82\linewidth]{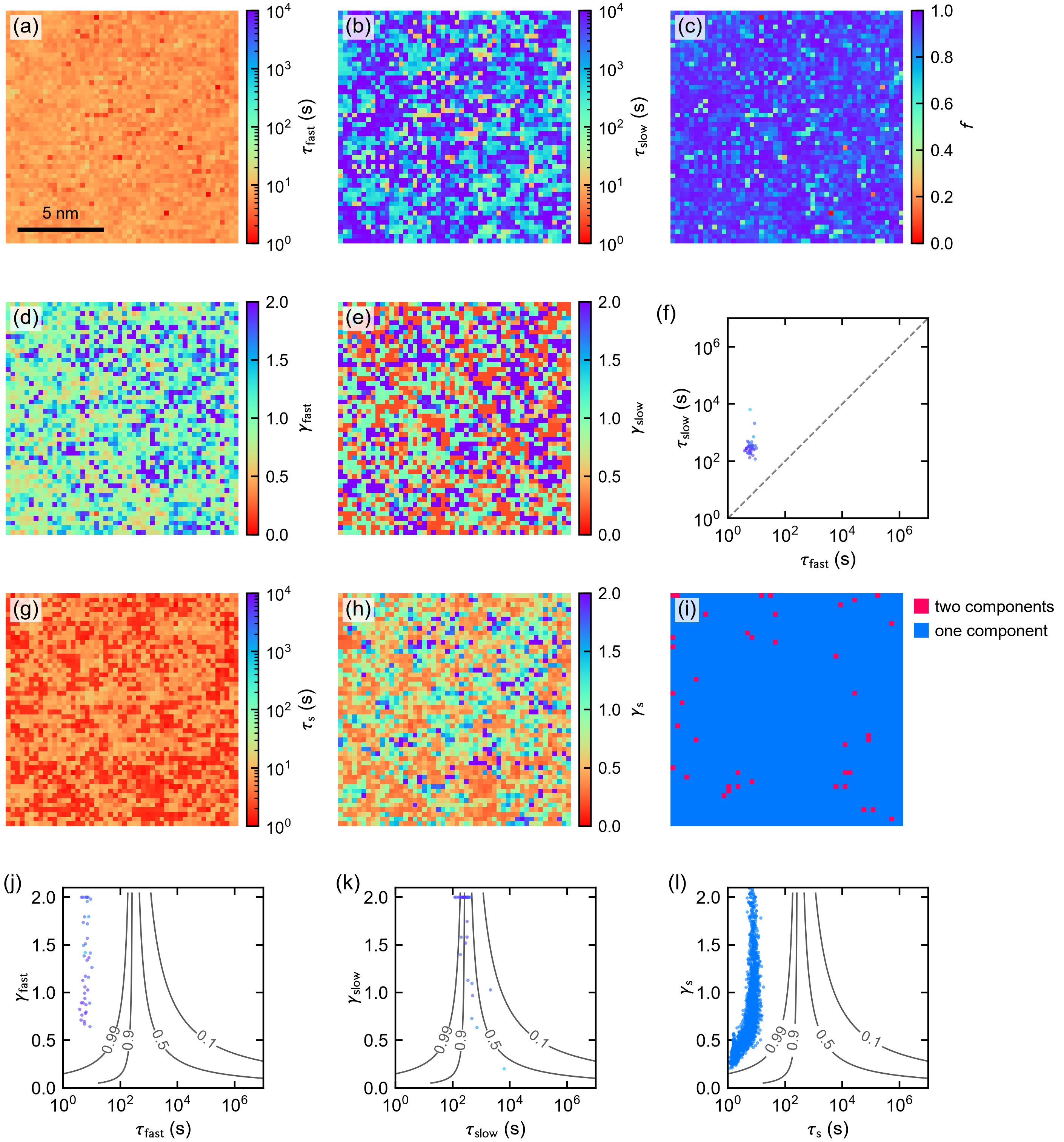}
\caption{Same as Fig.~\ref{fig:maps558} at 578~K.}
\label{fig:maps578}
\end{figure}

\begin{figure}[tbp]
\centering
\includegraphics[width=0.82\linewidth]{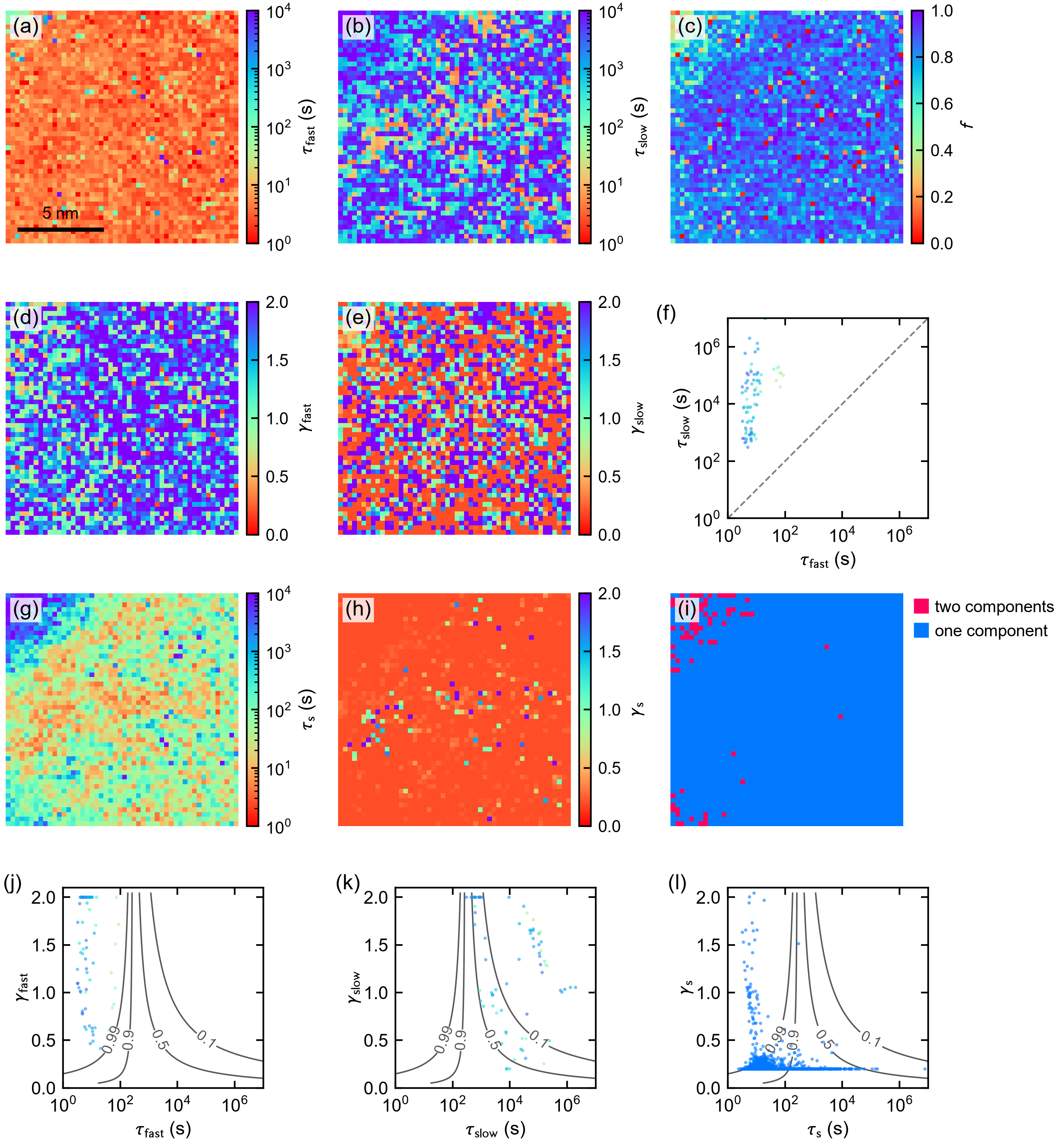}
\caption{Same as Fig.~\ref{fig:maps558} at 583~K.}
\label{fig:maps583}
\end{figure}

\begin{figure}[tbp]
\centering
\includegraphics[width=0.82\linewidth]{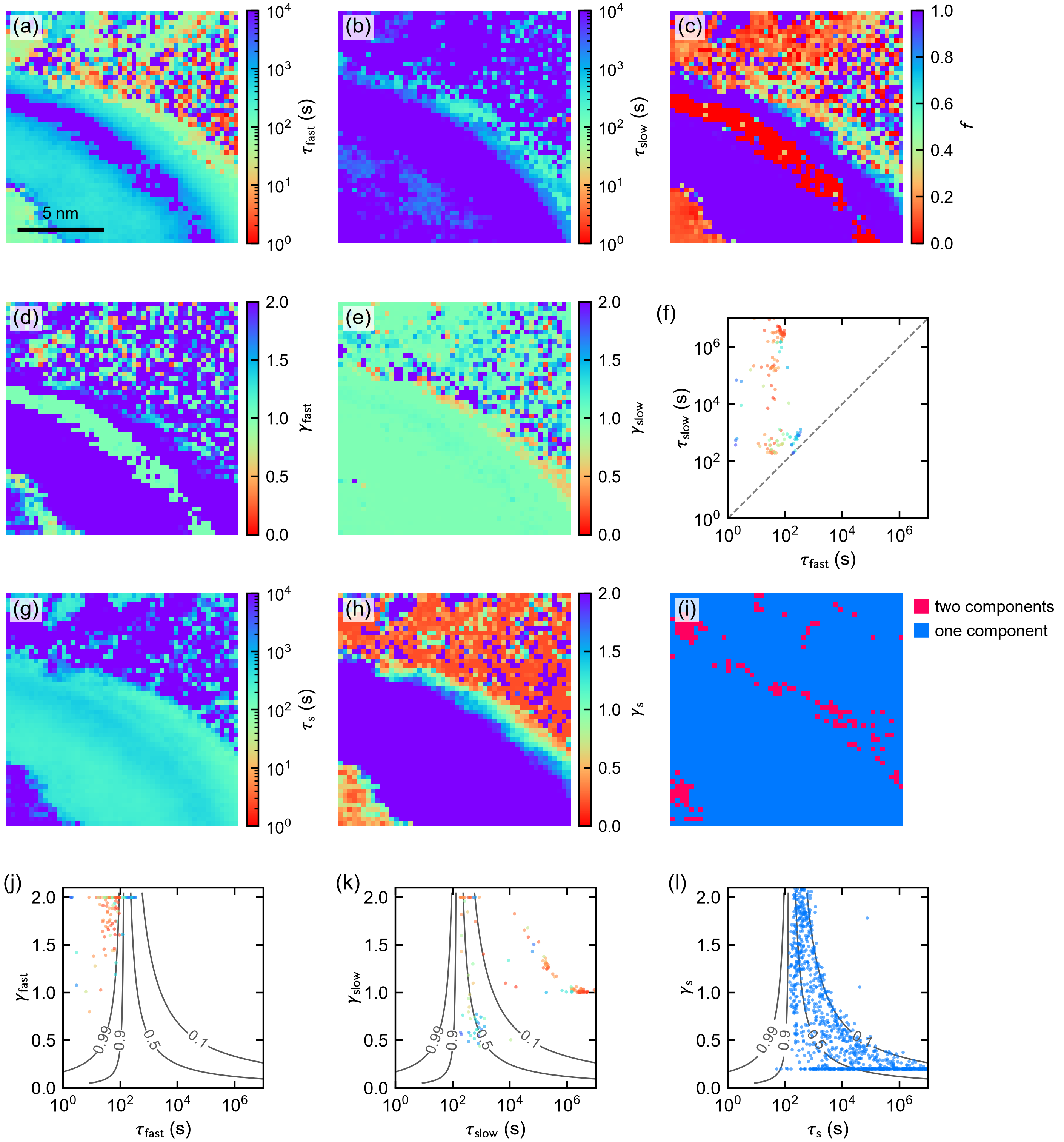}
\caption{Same as Fig.~\ref{fig:maps558} at 588~K.}
\label{fig:maps588}
\end{figure}

Figures~\ref{fig:maps558}--\ref{fig:maps588} show the maps of the relaxation parameters, the classification
of the positions by the BIC, and the relations between the parameters at each temperature. The fraction of
positions requiring two components is 0.61, 0.27, 0.11, 0.05, 0.02, 0.03, and 0.05 from 558 to 588~K. The data
at 583 and 588~K were analyzed in the same way only for this Supplement; at 588~K the positions requiring two components lie mainly along the boundary of the
crystallized region (Fig.~\ref{fig:maps588}).

\subsection{Structural parameters}

\begin{figure}[tbp]
\centering
\includegraphics[width=0.75\linewidth]{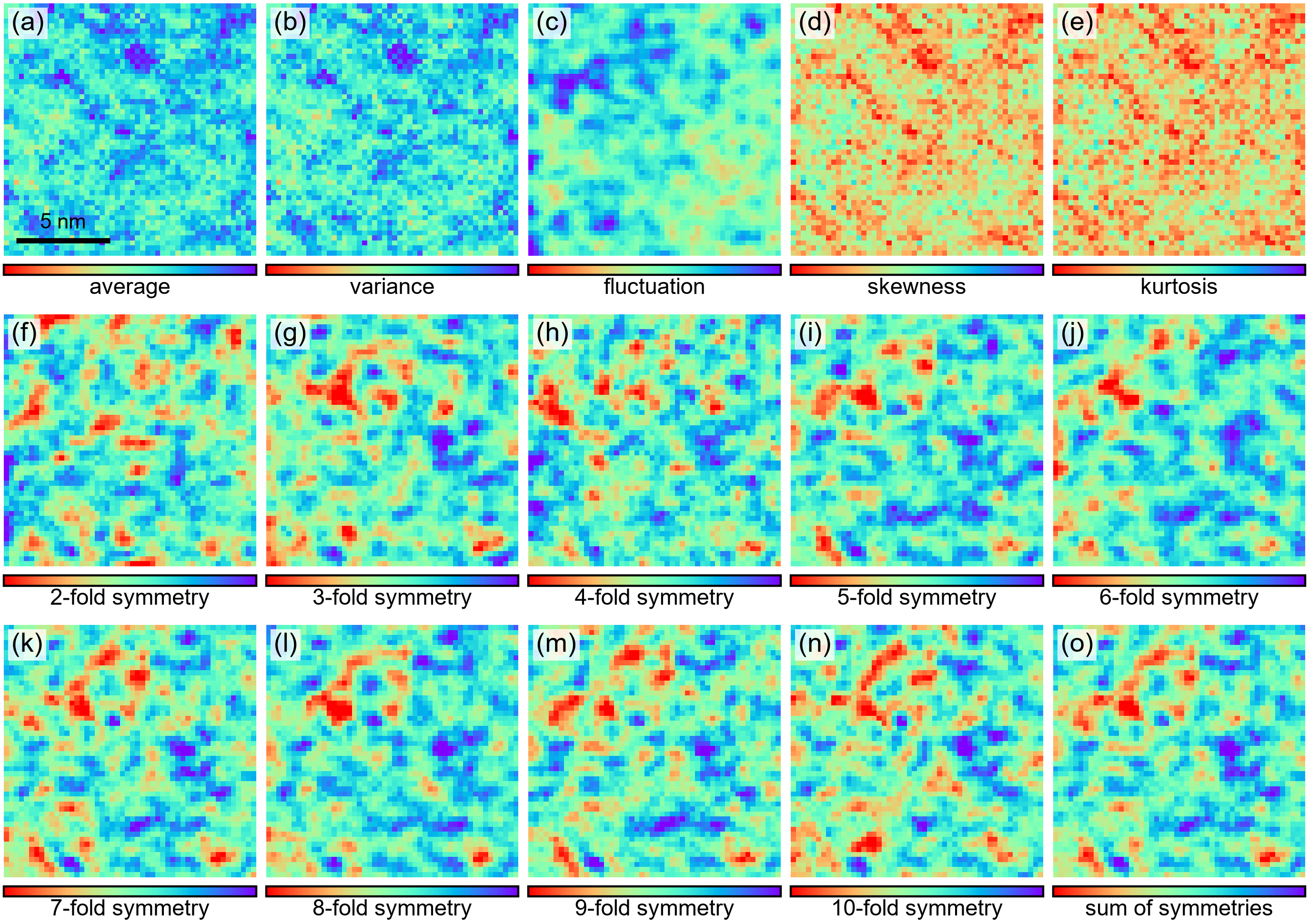}
\caption{Structural parameters at 558~K, averaged over the measurement: (a) average, (b) variance,
(c) fluctuation, (d) skewness, and (e) kurtosis of the diffracted intensity, (f)--(n) twofold to tenfold
rotational symmetries, and (o) the sum of the symmetries in (f)--(n). The color scales are common to all
temperatures.}
\label{fig:struct558}
\end{figure}

\begin{figure}[tbp]
\centering
\includegraphics[width=0.75\linewidth]{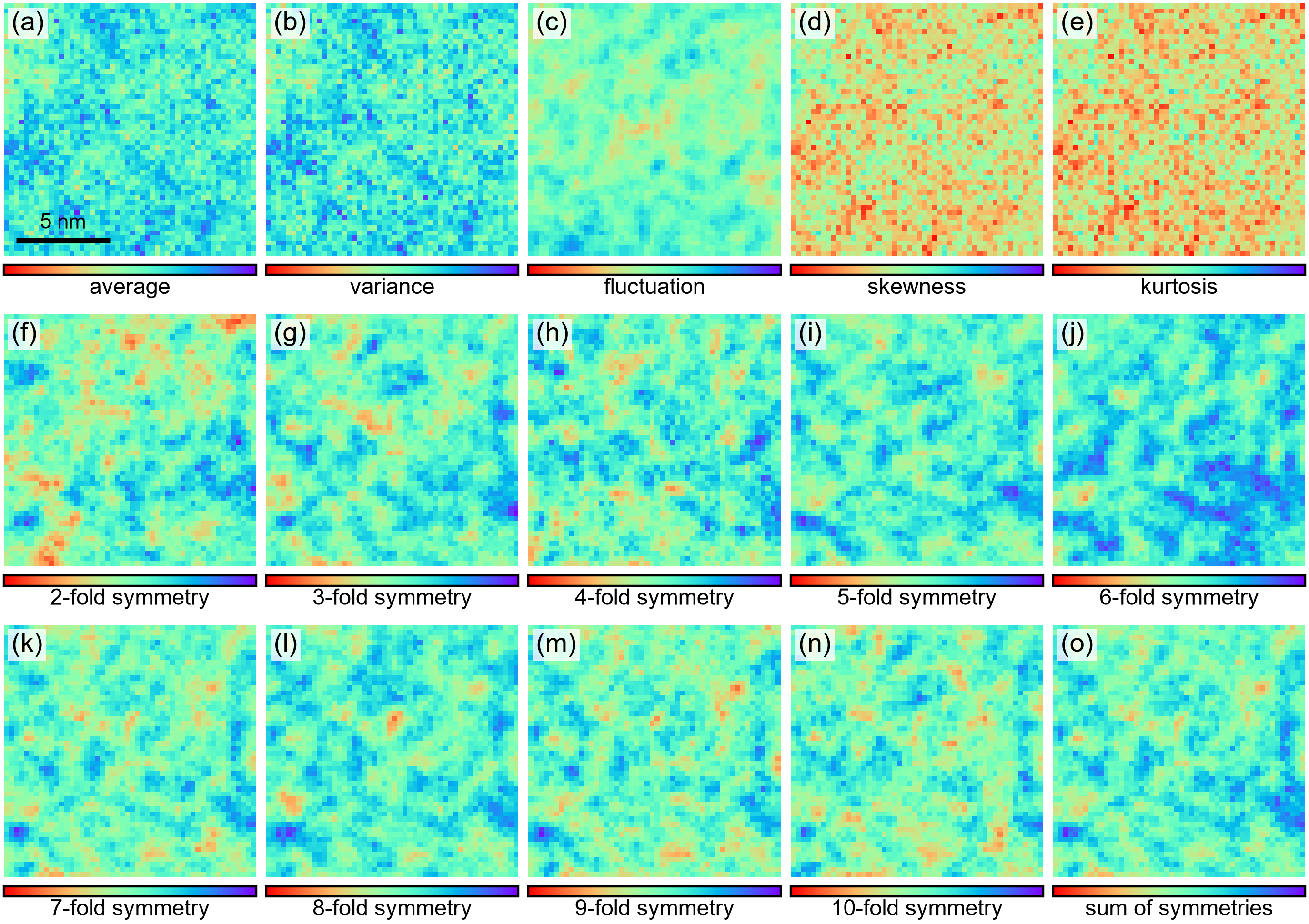}
\caption{Same as Fig.~\ref{fig:struct558} at 563~K.}
\label{fig:struct563}
\end{figure}

\begin{figure}[tbp]
\centering
\includegraphics[width=0.75\linewidth]{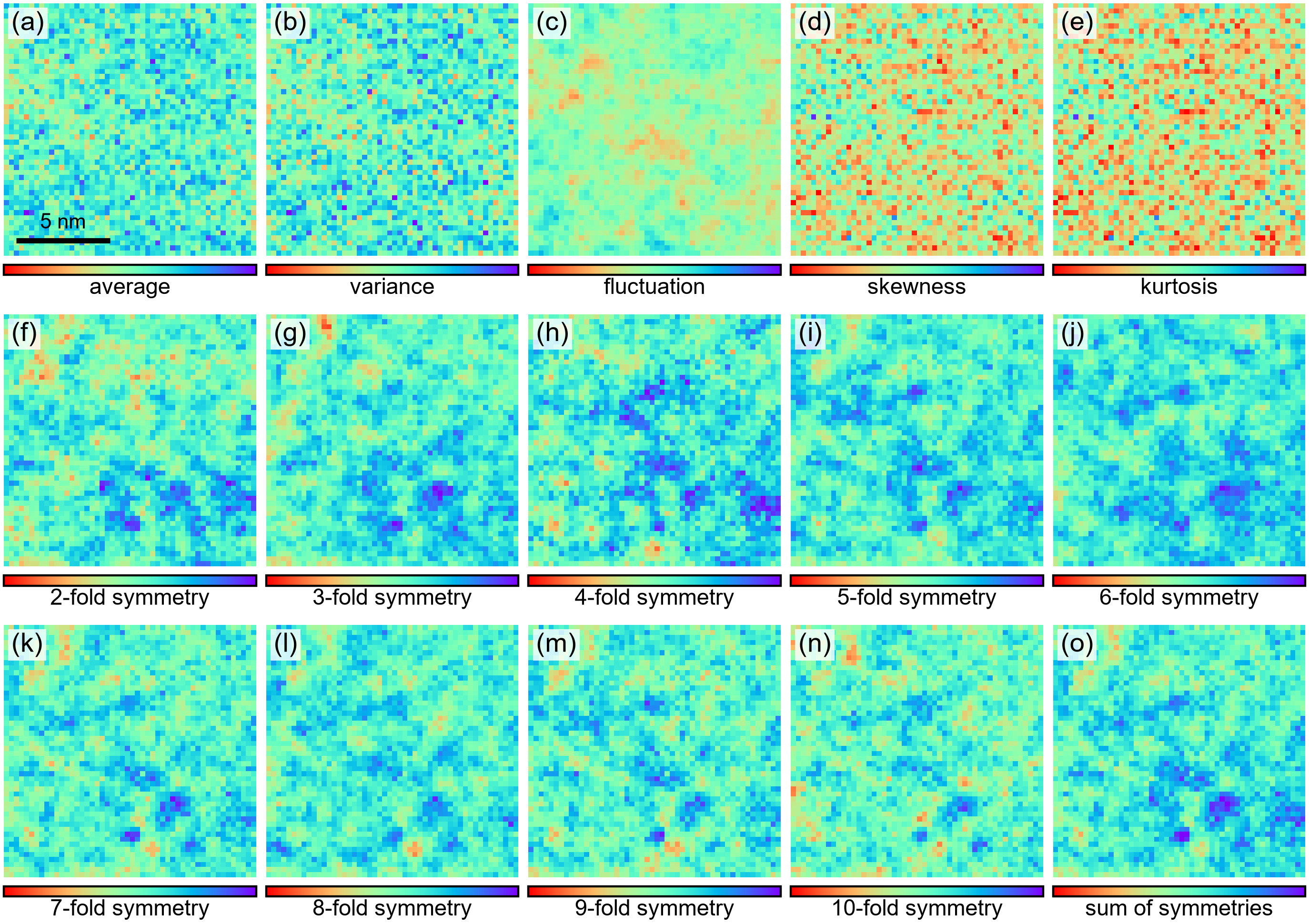}
\caption{Same as Fig.~\ref{fig:struct558} at 568~K.}
\label{fig:struct568}
\end{figure}

\begin{figure}[tbp]
\centering
\includegraphics[width=0.75\linewidth]{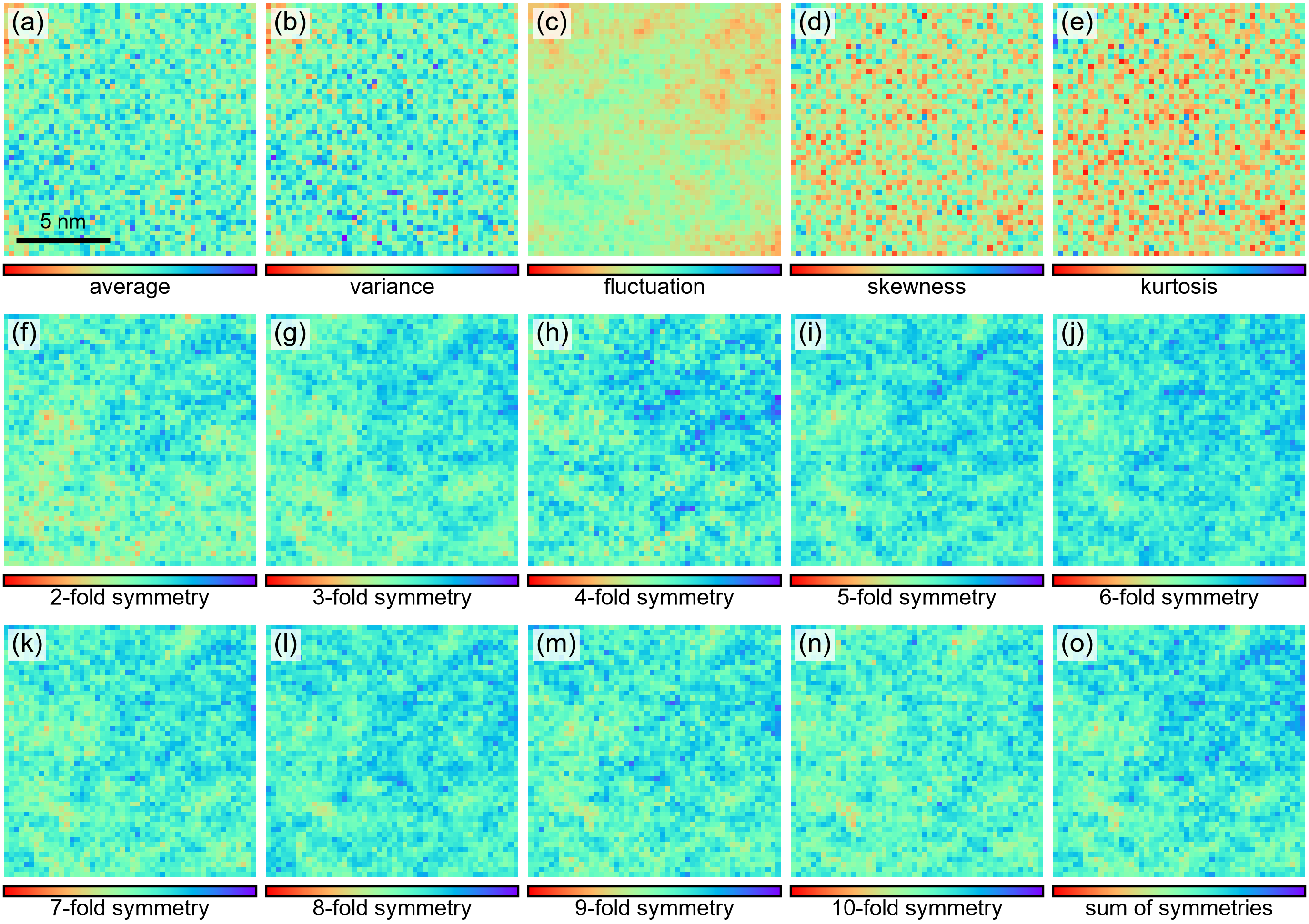}
\caption{Same as Fig.~\ref{fig:struct558} at 573~K.}
\label{fig:struct573}
\end{figure}

\begin{figure}[tbp]
\centering
\includegraphics[width=0.75\linewidth]{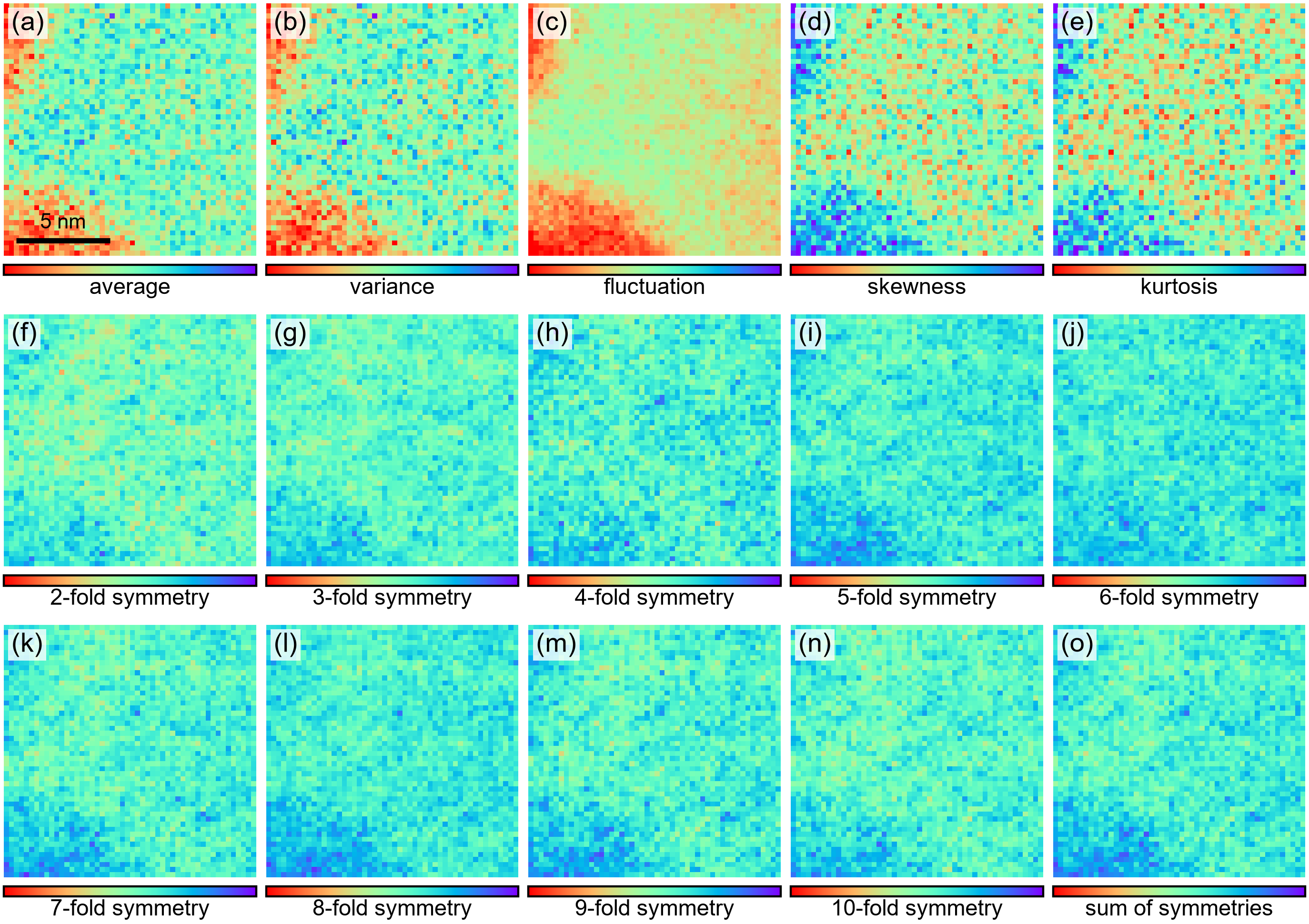}
\caption{Same as Fig.~\ref{fig:struct558} at 578~K.}
\label{fig:struct578}
\end{figure}

\begin{figure}[tbp]
\centering
\includegraphics[width=0.75\linewidth]{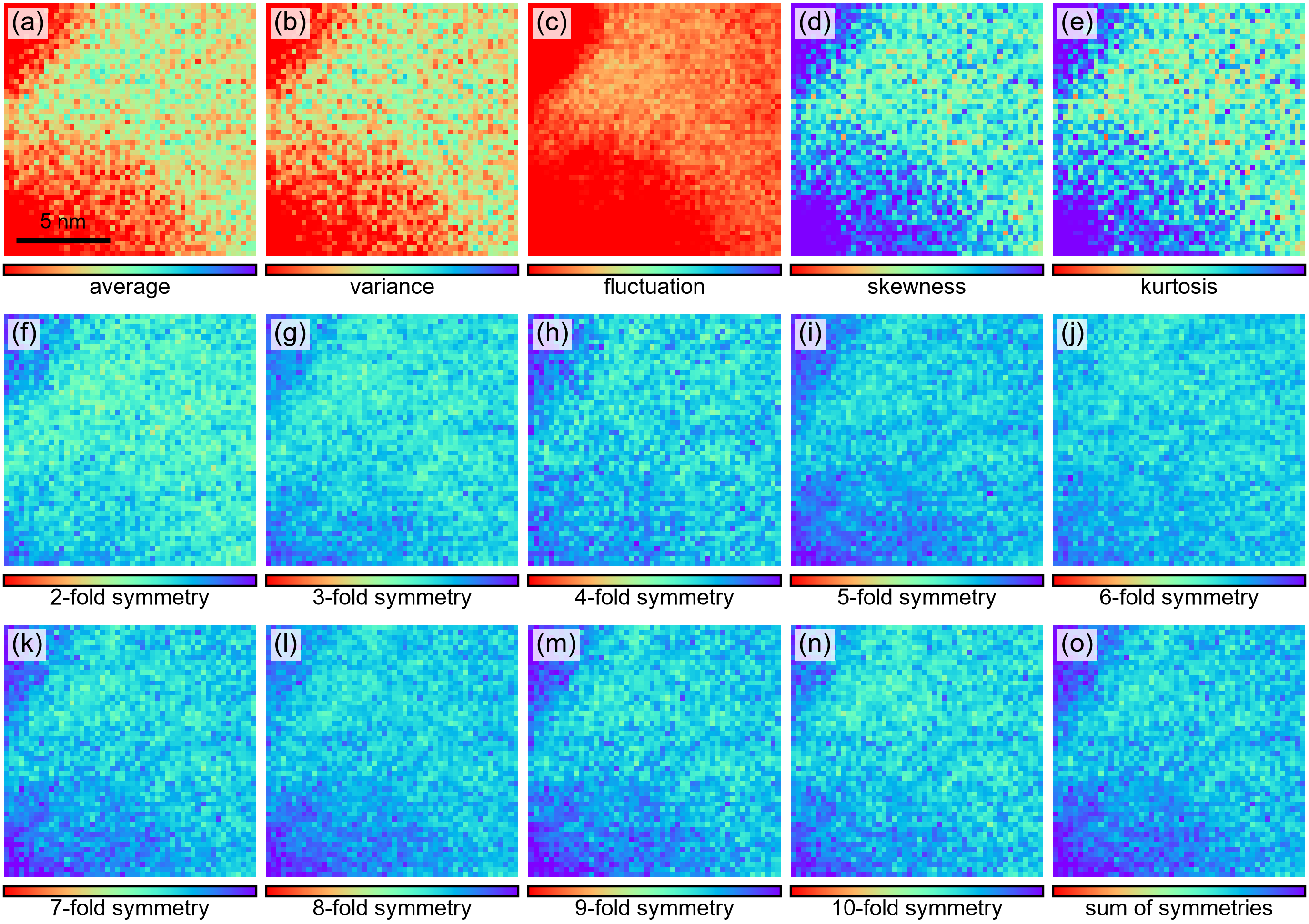}
\caption{Same as Fig.~\ref{fig:struct558} at 583~K.}
\label{fig:struct583}
\end{figure}

\begin{figure}[tbp]
\centering
\includegraphics[width=0.75\linewidth]{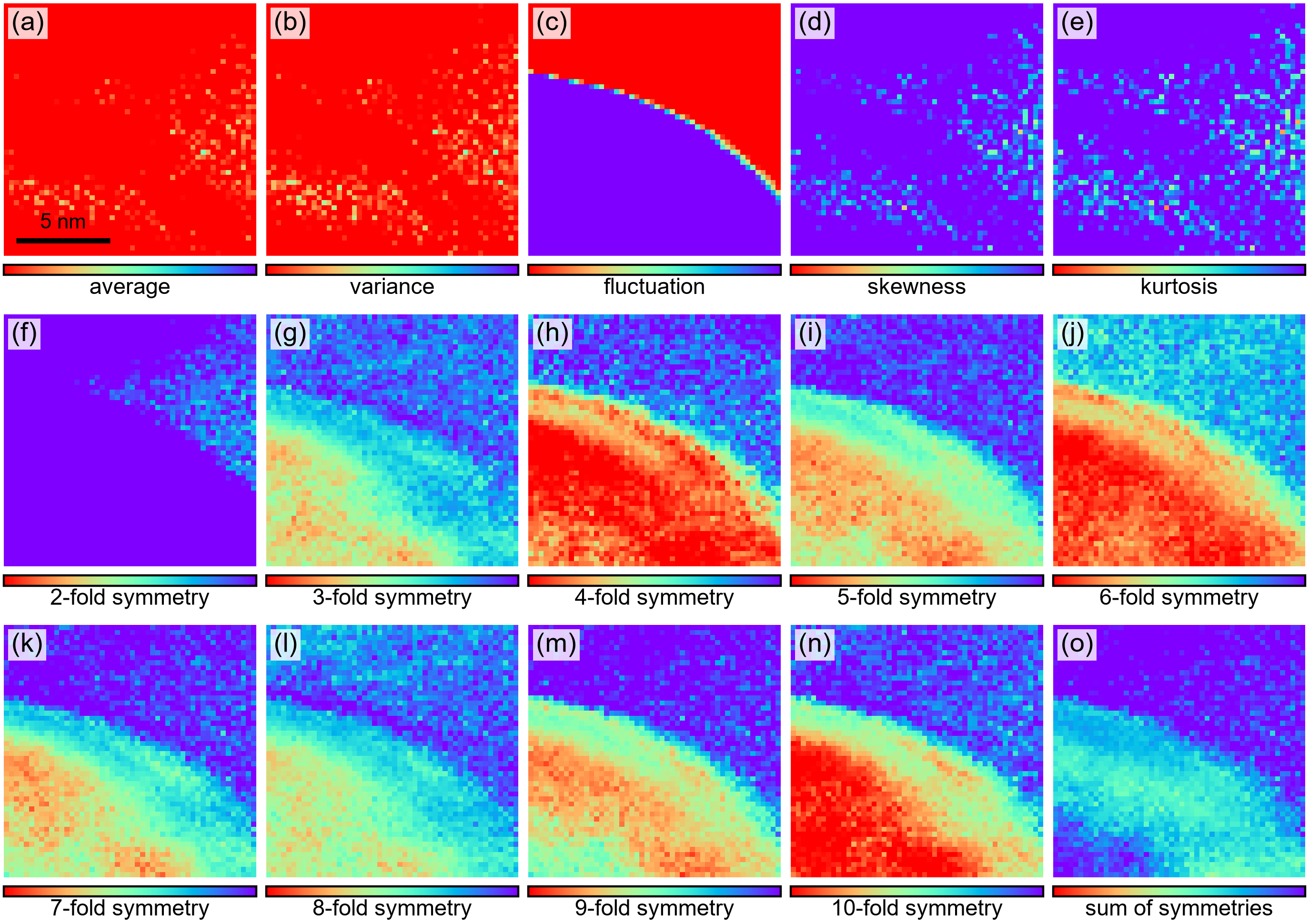}
\caption{Same as Fig.~\ref{fig:struct558} at 588~K.}
\label{fig:struct588}
\end{figure}

\begin{figure}[tbp]
\centering
\includegraphics[width=0.6\linewidth]{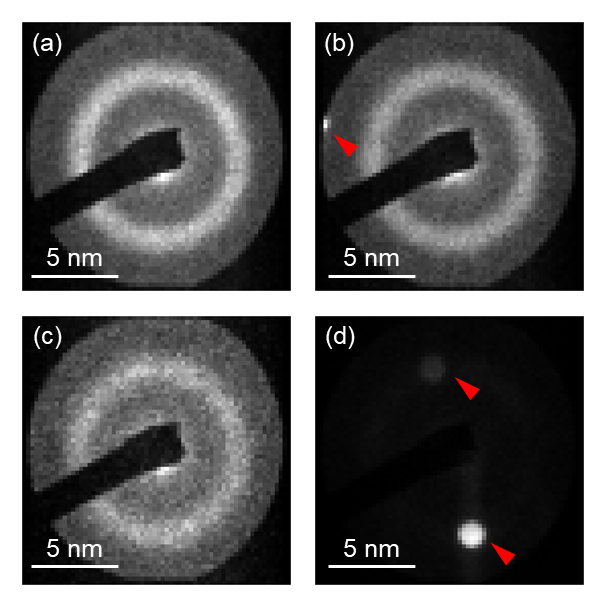}
\caption{Temporal averages of the diffraction patterns from regions of high and low fluctuation at 578
and 588~K. Arrows mark Bragg reflections from crystallites.}
\label{fig:cryst}
\end{figure}

Figures~\ref{fig:struct558}--\ref{fig:struct588} show the maps of the structural parameters averaged over
the measurement at each temperature. A common color scale is used for all temperatures. As the
temperature increases, the range of the values decreases and the maps become spatially more uniform.
This reflects the faster temporal change of the structure rather than a more uniform structure
(Sec.~\ref{sec:corr}, Fig.~\ref{fig:naccum}).
At 578~K the fluctuation is small at the upper left and lower left of the field of view
(Fig.~\ref{fig:struct578}), and a similar decrease on the left-hand side is seen at 583~K. These regions
have crystallized: the crystallites give Bragg reflections outside the range of spatial frequency used
for the analysis (Fig.~\ref{fig:cryst}), so the low values of the fluctuation there are an effect of
crystallization rather than a property of the glass. At 588~K the fluctuation and the twofold symmetry
increase at the bottom of the field of view, because Bragg reflections from crystallites now fall inside
the range of spatial frequency used for the analysis. For this reason, data above 578~K were not used
in the main text. At 578~K the crystallized part covers only about 7\% of the analyzed region
(Sec.~\ref{sec:corrB}); excluding it changes the fraction of two-component positions from 0.016 to 0.015,
the mean $f$ from 0.90 to 0.90, $\taus$ from 4.1 to 4.2~s, and the activation energy of $\taus$
from 604 to 603~kJ\,mol$^{-1}$.

\clearpage
\section{Correlations with the structural parameters before and after controlling for the fluctuation}
\label{sec:partial}

\begin{figure}[tbp]
\centering
\includegraphics[width=\linewidth]{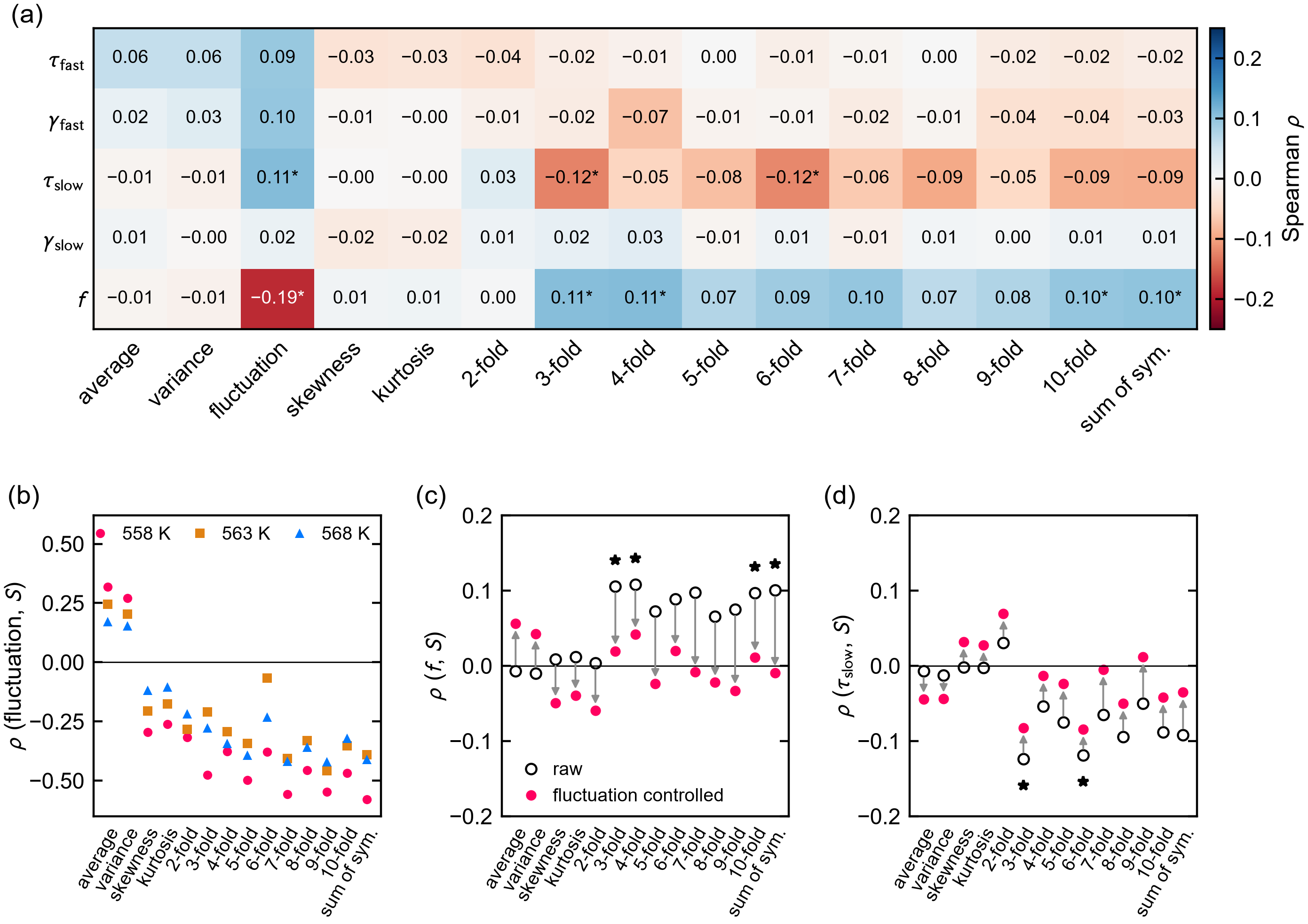}
\caption{Effect of controlling for the fluctuation $V$ on the correlations between the relaxation and the
structural parameters. (a) Spearman rank correlations at 558~K before the control; asterisks mark the
correlations that remain significant after the toroidal-shift surrogate test and the false-discovery-rate
correction. (b) Spearman rank correlation between $V$ and each of the other structural parameters at 558,
563 and 568~K. (c), (d) Correlations of (c) $f$ and (d) $\tauslow$ with each structural parameter at 558~K
before (open circles) and after (filled circles) $V$ is controlled for. Asterisks mark the correlations
that are significant before the control; none is significant after it.}
\label{fig:partial}
\end{figure}

Figure~\ref{fig:partial}(a) shows the Spearman rank correlations between the relaxation and the
structural parameters at 558~K before the fluctuation $V$ is controlled for. In addition to the
correlations of $f$ and $\tauslow$ with $V$, several $n$-fold symmetries appear to be correlated with the
relaxation: $f$ with the 3-, 4- and 10-fold symmetries and with their sum ($\rho = 0.10$--$0.11$), and
$\tauslow$ with the 3- and 6-fold symmetries ($\rho = -0.12$), all significant after the false-discovery-rate
correction.

These correlations are not independent of $V$. All structural parameters are correlated with $V$ across
positions [Fig.~\ref{fig:partial}(b)]: the average and the variance positively ($\rho = 0.15$--$0.32$ at
558--568~K), and the skewness, the kurtosis and all $n$-fold symmetries negatively (down to $\rho = -0.58$
for the sum of the symmetries at 558~K). Because $f$ is negatively and $\tauslow$ positively correlated
with $V$, a structural parameter that is negatively correlated with $V$ necessarily appears positively
correlated with $f$ and negatively correlated with $\tauslow$, even if it carries no information of its
own. The signs of all significant correlations with the $n$-fold symmetries in
Fig.~\ref{fig:partial}(a) are those expected from this mediation.

To remove the contribution of $V$, we computed the partial rank correlation
\begin{equation}
\rho_{RS\cdot V} = \frac{\rho_{RS} - \rho_{RV}\,\rho_{SV}}{\sqrt{(1-\rho_{RV}^{2})(1-\rho_{SV}^{2})}},
\end{equation}
where $R$ is a relaxation parameter, $S$ a structural parameter, and the $\rho$ are Pearson correlations
of the ranks. Its significance was assessed with the same toroidal-shift surrogate test, in which the map
of $S$ is shifted cyclically, followed by the false-discovery-rate correction at each temperature.
Figures~\ref{fig:partial}(c) and (d) show how the correlations of $f$ and $\tauslow$ change: the
correlations of $f$ with the 3-, 4- and 10-fold symmetries fall from 0.10--0.11 to 0.01--0.04, and those
of $\tauslow$ with the 3- and 6-fold symmetries from $-0.12$ to $-0.08$. None of them remains significant.
The same holds at 563 and 568~K: after $V$ is controlled for, no structural parameter other than $V$
shows a significant correlation with any relaxation parameter ($|\rho| \le 0.13$). The apparent
correlations with the $n$-fold symmetries are therefore copies of the correlation with $V$, and the
$n$-fold symmetries carry no independent structural information on the relaxation. The partial
correlations at 558~K are shown in Fig.~6 of the main text.

\clearpage
\bibliography{refs}